\documentclass[12pt,preprint]{aastex}

\shorttitle{The opaque nascent starburst in NGC\,1377}
\shortauthors{Roussel et al.}

\begin{document}

\title{The opaque nascent starburst in NGC\,1377: Spitzer SINGS observations}

\author{H. Roussel\altaffilmark{1,2},
G. Helou\altaffilmark{1},
J.D. Smith\altaffilmark{3},
B.T. Draine\altaffilmark{4},
D.J. Hollenbach\altaffilmark{5},
J. Moustakas\altaffilmark{3},
H.W. Spoon\altaffilmark{6},
R.C. Kennicutt\altaffilmark{7,3},
G.H. Rieke\altaffilmark{3},
F. Walter\altaffilmark{2},
L. Armus\altaffilmark{1},
D.A. Dale\altaffilmark{8},
K. Sheth\altaffilmark{1},
G.J. Bendo\altaffilmark{9},
C.W. Engelbracht\altaffilmark{3},
K.D. Gordon\altaffilmark{3},
M.J. Meyer\altaffilmark{10},
M.W. Regan\altaffilmark{10},
E.J. Murphy\altaffilmark{11}
}

\affil{1) California Institute of Technology, Pasadena, CA 91125}
\affil{2) Max-Planck-Institut f\"ur Astronomie, Heidelberg, 69117 Germany}
\affil{3) Steward Observatory, University of Arizona, Tucson, AZ 85721}
\affil{4) Princeton University, Princeton, NJ 08544}
\affil{5) NASA Ames Research Center, Moffett Field, CA 94035}
\affil{6) Cornell University, Ithaca, NY 14853}
\affil{7) IoA, University of Cambridge, UK}
\affil{8) University of Wyoming, Laramie, WY 82071}
\affil{9) Imperial College, London SW7 2AZ, UK}
\affil{10) Space Telescope Science Institute, Baltimore, MD21218}
\affil{11) Yale University, New Haven, CT 06520}
\email{roussel@mpia-hd.mpg.de}

\begin {abstract}
We analyze extensive data on NGC\,1377 from the Spitzer Infrared Nearby Galaxies
Survey (SINGS). Within the category of nascent starbursts, that we previously
selected as having infrared to radio continuum ratios in large excess of the
average, and containing hot dust, NGC\,1377 has the largest infrared excess
yet measured. Optical imaging reveals a morphological distortion suggestive of
a recent accretion event. Infrared spectroscopy reveals a compact and opaque
source dominated by a hot, self-absorbed continuum ($\tau \sim 20$ in the
10\,$\mu$m silicate band). We provide physical evidence against non-stellar
activity being the heating source.
H{\small II} regions are detected through the single [Ne{\small II}] line,
probing $< 1$\% of the ionizing radiation. Not only is the optical depth in
different gas and dust phases very high, but $> 85$\% of ionizing photons are
suppressed by dust. The only other detected emission features are molecular
hydrogen lines, arguably excited mainly by shocks, besides photodissociation
regions, and weak aromatic bands.
The new observations support our interpretation in terms of an extremely young
starburst ($ <1$\,Myr). More generally, galaxies deficient in radio synchrotron
are likely observed within a few Myr of the onset of a starburst and after a
long quiescence, prior to the replenishment of the interstellar medium
with cosmic rays.
The similar infrared-radio properties of NGC\,1377 and some infrared-luminous
galaxies suggest that NGC\,1377 constitutes an archetype to better understand
starburst evolution. Although rare locally because observed in a brief evolutionary
stage, nascent starbursts may represent a non-negligible fraction of merger-induced
starbursts dominating deep infrared counts.
Since they differ dramatically from usual starburst templates, they bear important
consequences for the interpretation of deep surveys.
\end {abstract}
 

\section{Introduction}

The infrared-radio correlation of star-forming galaxies \citep{Helou85, Jong85}
suffers very few exceptions. At centimeter wavelengths, the radio continuum
of massive galaxies is dominated by synchrotron emission, from cosmic ray
electrons previously accelerated in type-II supernova remnants, propagating in
the interstellar magnetic field, and decaying in less than $10^8$\,years
\citep{Condon92}. The emission from thermal electrons in H{\small II} regions
usually contributes only of the order of 20\% of the total emission of galaxies
at 3\,cm, and of the order of 5\% at 20\,cm. The far-infrared continuum measures
the energy peak of the dust emission, and is a fair tracer of the instantaneous
star formation rate for dusty starburst galaxies. In more quiescent galaxies,
a significant fraction of the far-infrared emission can originate from dust heated
by intermediate-mass stars \citep{Lonsdale87, Sauvage92}.

NGC\,1377 is the most extreme example known so far of the rare category of
non-dwarf galaxies which have infrared-to-radio ratios in significant excess
of the average, and where the infrared excess is not attributable to cirrus-like
emission~; this is because
the high dust temperatures of the selected galaxies imply high
radiation field intensities
\citep{Roussel03, Roussel05}. In NGC\,1377, synchrotron emission is deficient by 
at least a factor 37 (i.e. by $> 8 \sigma$) with respect to normal galaxies.
In addition, H{\small II} regions were not detected through near-infrared hydrogen
recombination lines nor through the thermal radio continuum, and the upper limits
that we obtained imply that they are fainter than expected for normal starbursts
by at least 70\%, likely due to the fact that most ionizing photons are absorbed
by dust.

The infrared and radio emission both have their source in the star formation
activity. Given the lack of direct connection between the physical processes
regulating both types of emission, however, the infrared-radio correlation can
hold only if strong coupling mechanisms operate \citep{Beck88, Helou93, Niklas97}.
First, the production rate of cosmic rays has to be roughly proportional to
that of dust-heating photons (which is achieved for a constant initial mass
function). In addition,
the magnetic field intensity has to be coupled with either the gas density
or the star formation rate (e.g. turbulence driven by massive star formation
inducing rapid amplification of the magnetic field). Empirically, starbursts
seem to constantly adjust their magnetic field so that the ratio of magnetic
energy density to radiation energy density remains constant, including
ultraluminous galaxies with very intense radiation fields
\citep{Condon91, Lisenfeld96}.

The infrared emission and the radio continuum both persist for long periods
subsequent to an instantaneous episode of star formation, but follow very
different timescales. In particular, cosmic rays are accelerated only about
4\,Myr after the birth of massive stars evolving into type-II supernovae,
whereas dust is heated almost instantaneously. This time delay implies that
the relative amounts of infrared and radio emission are expected to vary
significantly, which can in principle be exploited to constrain the age of
individual star forming regions, i.e. regions with relatively simple star
formation histories. Galaxies, however, host young stellar populations which
span an extended age range, and star formation generally goes on for long
periods, either continuously or in the form of recurrent bursts. For large
regions within galaxies, systematic variations of the infrared-to-radio
ratio are thus in practice erased by complex star formation histories.
Non-uniform interstellar medium properties (structure and dust content)
will also induce further variations decoupled from age.

Discrete sources with radio spectra dominated by optically-thick thermal emission,
from ultracompact H{\small II} regions, have been found and studied at high
angular resolution within starbursting galaxies, most of them dwarfs: e.g. by
\citet{Turner98} in NGC\,5253, \citet{Kobulnicky99} in He\,2-10, \citet{Tarchi00}
in NGC\,2146, \citet{Beck00} in several Wolf-Rayet galaxies, \citet{Beck02} in
II\,Zw\,40. These systems thus contain a number of very young star forming
sites. On global scales, the radio continuum of blue compact dwarf galaxies can
be dominated by the thermal component.
In such objects, both the radio and
the infrared emission deviate significantly from the scaling relations with star
formation rate followed by more massive galaxies
\citep[][and references therein]{Hunt05}~: the radio continuum is depleted because
of the delayed injection of cosmic rays in a young starburst, and possibly because
of enhanced cosmic ray escape~; and the infrared emission is depleted because of
lower dust content, different dust properties, and higher porosity of the
interstellar medium.

We selected the galaxies that we term nascent starbursts from an infrared
flux-limited sample by their high infrared to radio flux ratios and their
high dust temperatures \citep{Roussel05}.
These criteria are designed to select galaxies hosting very young starbursts,
like the galaxies mentioned above, but constituting a very different population:
their infrared luminosities are much higher, they are more massive, and if cosmic
ray escape is less efficient than in dwarfs, the radio deficiency also implies
that cosmic rays produced by past star formation episodes have decayed, and thus
have been injected more than $\sim 100$\,Myr ago. In other words, the intensity
contrast between the current starburst and the past star formation integrated over
several tens of million years is very high. While classical starburst galaxies also
contain very young star forming sites, our view of these is confused by recurrent
burst episodes or continuous star formation in the recent past, which produce average
signatures typical of evolved bursts, mimicking single populations of ages
$\sim 5$--10\,Myr \citep[e.g.][]{Calzetti97}.
Nascent starbursts, on the other hand, offer an ideally simplified view because
of the dominance of a single and quasi-instantaneous burst of star formation
(previous stellar generations being too old to contribute to the production of
star formation tracers).

The nascent starbursts are extremely rare. They represent a negligible fraction
($\approx 1$\%) of an infrared flux-limited sample in the local universe
such as the IRAS Faint Galaxy Sample, and about 16\% of galactic systems
whose flux density is higher at 60\,$\mu$m than at 100\,$\mu$m
\citep{Roussel03}. These objects nevertheless deserve careful study, because
they constitute ideal settings to better understand the initial conditions and
early evolution of starbursts, as well as the regulation of the infrared-radio
correlation.

In such objects, where the active regions are very young and compact, the
most useful diagnostics on the nature of the heating source are expected to
arise from the dust and the molecular gas phases, because ultraviolet radiation
will be confined to dense ultracompact H{\small II} regions \citep{Habing79}.
The mid-infrared spectral range is of particular interest not only to constrain
the nature and excitation of the small-size dust grains and aromatic compounds,
but also because it contains a series of forbidden and molecular hydrogen lines
providing unique diagnostics on H{\small II} regions, photodissociation
regions, and the outer layers of molecular clouds, and which are much less
affected by extinction than the other diagnostics accessible in the optical
and near-infrared domains. An ISOPHOT mid-infrared spectrum of NGC\,1377
between 2.5 and 4.8\,$\mu$m and between 6 and 11.5\,$\mu$m was presented
by \citet{Laureijs00}, as part of a small sample of galaxies whose infrared
flux density distribution peaks at 60\,$\mu$m. This spectrum appeared very
peculiar in comparison with normal galaxies, containing a broad emission
feature between 6 and 8.5\,$\mu$m in place of aromatic bands universally
found in metal-rich star-forming galaxies, but lacked sufficient wavelength
coverage to draw conclusions regarding the nature of the source and the
excitation mechanisms of the dust and gas. NGC\,1377 was selected to be part
of the Spitzer Infrared Nearby Galaxies Survey (SINGS) sample \citep{Kennicutt03}
to extend the range of explored $F_{60\,\mu m}/F_{100\,\mu m}$ ratios, indicative
of average dust temperatures. In this paper, we present the SINGS broadband
imaging data (3 to 160\,$\mu$m) and spectral imaging data (with continuous
coverage between 5 and 38\,$\mu$m) of NGC\,1377, as well as ancillary optical
imaging and spectroscopic data obtained for SINGS.
We analyze them in view of
constraining the geometry of the gas and dust reservoir, the excitation of the
different phases of the interstellar medium probed in the 5--38\,$\mu$m range,
and what these properties imply for the nature of the activity in NGC\,1377.
The hypothesis of non-stellar activity is discussed in Section~\ref{seyfert}.

\section{SINGS data}
\label{data}

\subsection{Spitzer spectral maps}
\label{spectra}

Low spectral resolution cubes ($\lambda / \Delta \lambda \approx 60$--130)
between 5 and 38\,$\mu$m and high spectral resolution cubes
($\lambda / \Delta \lambda \approx 600$) between 10 and 37\,$\mu$m centered
on the nucleus of NGC\,1377 were acquired with the IRS instrument
\citep{Houck04a} in February 2005 and August 2004, respectively. For each
spectral resolution mode, the full wavelength range is covered in two parts
by two different slits: at low resolution, SL (5--14.5\,$\mu$m) and
LL (14--38\,$\mu$m); and, at high resolution, SH (10--19.5\,$\mu$m) and
LH (19--37\,$\mu$m). Quadruple coverage is obtained at each point of the maps,
except at the edges, by moving the slit in increments of half the slit size
in each direction (see \citet{Kennicutt03} and \citet{Smith04} for details
on the observing strategy in SINGS). Given that the infrared emission of
NGC\,1377 is not spatially resolved (Section~\ref{morph}), all the emission is
recovered within the spectral maps, the smallest of which, in the SH module,
is $23\arcsec \times 15$\arcsec. The data were pre-processed with the S12
version of the Spitzer Science Center pipeline, and then assembled in
spectral cubes with Cubism.
Cubism is the software
developed to rectify, project
and co-add SINGS data, with the capability to subtract a spectral
cube of the background and foreground (zodiacal) emission extracted from the
data themselves, to mask pixels exhibiting an abnormal responsivity, and to
perform spatial and spectral extractions (Smith et al., 2006, in preparation).
In order to improve the signal to noise ratio of the high-resolution spectral
data, an extra step was included in the processing of these data, outside of
Cubism. In the SH and LH modules, the slit is very narrow, only 6 pixels-wide on
average, and the pixels with an abnormal responsivity are more numerous than in
the SL and LL modules. Because of the minimal redundancy, masking pixels can
create spurious features in the final cube. To mitigate this effect, the stacked
diffraction images were inspected for obvious deviant pixels (i.e. features that
are narrower than the spectral and spatial resolution), and these pixels were
interpolated in each pointing along the wavelength dimension. This treatment is
appropriate for infrared point sources like NGC\,1377, with abrupt spatial variations
and a smoothly-varying continuum, and leaves intact real spectral features.

\subsection{Spitzer broadband imaging}
\label{ir_images}

Images in the four bands of the IRAC camera \citep{Fazio04}, centered at
effective wavelengths of 3.6, 4.5, 5.7 and 7.9\,$\mu$m (usually designated by
their fiducial wavelengths 3.6, 4.5, 5.8 and 8.0\,$\mu$m), were obtained in
August 2004. Scan maps in the three bands of the MIPS instrument
\citep{Rieke04} at effective wavelengths of 24, 71 and 156\,$\mu$m (fiducial
wavelengths of 24, 70 and 160\,$\mu$m)
were acquired in September 2004. The observing strategy and data reduction are
described by \citet{Kennicutt03}. The full width at half maximum of the
point spread function (PSF) is close to 2.0\arcsec\ at 7.9\,$\mu$m, or 200\,pc
at the distance of NGC\,1377 (21\,Mpc).
Flux calibration uncertainties are of the order of 10\% in the IRAC bands, and
5\%, 10\% and 15\% in the MIPS 24, 71 and 156\,$\mu$m bands.

Broadband infrared flux densities of NGC\,1377 are listed in Table~\ref{tab_flux}.
In the IRAC short-wavelength bands, both the total fluxes and the nuclear fluxes
are provided. In all IRAC bands, the nuclear fluxes were measured within the
smallest point-source aperture of diameter 4.88\arcsec, with a background annulus
of internal and external diameters of 4.88\arcsec\ and 14.64\arcsec, in which
case aperture correction factors ranging from 1.213 to 1.584 are necessary
(IRAC handbook). Use of the other point-source apertures tabulated in the
IRAC handbook, for the 5.7\,$\mu$m and 7.9\,$\mu$m bands, leads to fluxes
higher by up to 8\% and 15\% respectively. The smallest aperture is preferable
in our case to minimize contamination by ghosts and other artifacts (see
Section~\ref{morph}).
It is also preferable, for the 3.6\,$\mu$m and 4.5\,$\mu$m bands,
because the underlying disk emission is non-negligible at these wavelengths
(see Table~\ref{tab_flux}). No extended emission is detected in the 5.7
and 7.9\,$\mu$m bands, and total fluxes measured in a large aperture (with
correction factors of 0.63 and 0.69, respectively, that are appropriate for
extended emission) are lower than the nuclear fluxes by 25\% and 6\%
respectively. Since the aperture correction factors for extended emission
are much more uncertain than for point sources, we adopt the nuclear fluxes as
total measurements in the 5.7 and 7.9\,$\mu$m bands. For the total fluxes in
the 3.6 and 4.5\,$\mu$m bands, we applied correction factors of 0.94 (IRAC
handbook).
As the MIPS detectors do not suffer from the light scattering phenomenon making
accurate photometry difficult in IRAC images, fluxes in the MIPS bands were simply
measured by summation over the whole extent of the source.

\subsection{Optical imaging and spectroscopy}
\label{opt_images}

Images in the B, V, R and I bandpasses were acquired at the CTIO 1.5\,m
telescope with the CFCCD (Cassegrain Focus CCD) instrument, in October
2001. The pixel size is 0.43\arcsec\ and the angular resolution (full
width at half maximum), measured on stars nearby NGC\,1377, varies
between 1.15\arcsec\ and 1.23\arcsec\ in the different bandpasses.
For comparison between the B band and the I band, the images were
registered using the positions of a dozen stars, with a residual radial
dispersion of 0.2\arcsec, and convolved to a common resolution of 1.5\arcsec.

Optical long-slit spectra between 3600 and 6900\AA, at a spectral resolution
of 8\AA, were acquired at the Steward Observatory 2.3\,m telescope in November 2001.
The observations and data reduction are described by Moustakas et al. (2006,
in preparation). We used an extraction aperture of $2.5\arcsec \times 10\arcsec$
centered on the nucleus of the galaxy and oriented along the minor axis.

\subsection{Flux calibration of the infrared spectra}
\label{calib}

The default flux calibration of Spitzer spectral data is appropriate only
for staring observations of point sources, in which the loss of flux due to
the diffraction beam being larger than the slit width is indirectly corrected for,
by matching staring observations of stars with spectrophotometric models.
Additional steps are necessary to calibrate mapping observations of
arbitrarily-shaped sources. Although NGC\,1377 is a point source in the infrared
(Sect.~\ref{morph}), we cannot use the default flux calibration because it was
not perfectly centered in the slits.
The flux loss fraction as a function of wavelength was derived from estimates
of the beam profile using the PSF simulator distributed by the Spitzer Science
Center.\footnote{http://ssc.spitzer.caltech.edu/archanaly/contributed/browse.html}
The spectra of the galaxy were then corrected by this function to remove the
added point-source flux falling outside the central slit, because this flux
is already retrieved by mapping the source. For the low-resolution data,
it is also necessary to correct for the fact that the default flux calibration
is done by extracting stellar spectra within an aperture much smaller than the
slit length. The conversion from e$^-$/s units to flux density units using the full
slit length was derived from highly redundant observations of the calibrating stars
HR\,6688 (for the LL module) and HR\,7891 (for the SL module), mapped in the
slit-parallel direction. The spectra of these stars, extracted in an aperture
of the same width as that used for NGC\,1377, were divided by model spectra
from \citet{Decin04}\footnote{The data are made available at the web address:
http://ssc.spitzer.caltech.edu/irs/calib/templ/ .} to derive the flux calibration
functions. These functions were adjusted with smoothly-varying polynomials before
being used to divide the LL and SL spectra of NGC\,1377. Broadband photometry in
the IRAC, MIPS and ISOCAM filters was then simulated from the spectra, using the
appropriate transmission curves, and compared with imaging photometry. We obtain
excellent agreement at 24\,$\mu$m and 5.7\,$\mu$m:
$F_{24}{\rm (MIPS)} / F_{24}{\rm (LL)} = 1.01$ and
$F_{5.7}{\rm (IRAC)} / F_{5.7}{\rm (SL)} = 0.98$. The agreement with ISOCAM data
is also quite good despite uncertainties in the saturation correction in the images
\citep[for details, see][]{Roussel03}:
$F_{6.7}{\rm (ISOCAM)} / F_{6.7}{\rm (SL)} = 0.88$ and
$F_{15}{\rm (ISOCAM)} / F_{15}{\rm (SL-LL)} = 0.97$. Note that the spectral shape
between 13 and 14.5\,$\mu$m is not well calibrated at low resolution, because of an
artifact at the 10\% level in the detector, which cannot be modelled at present.
Comparison with SH
data, which are not affected by this artifact, however shows good consistency.
We also obtain $F_{7.9}{\rm (IRAC)} / F_{7.9}{\rm (SL)} = 0.92$. Since IRAC photometry
is itself uncertain (see Sect.~\ref{ir_images}), this does not imply any significant
error in the spectral flux calibration.

\section{Morphology and general properties}
\label{morph}

NGC\,1377 is a member of the unrelaxed Eridanus galaxy group \citep{Willmer89},
at an estimated distance of 21\,Mpc. In stellar light, it has the
appearance of a regular lenticular galaxy, with a diameter of 1.8\arcmin\
(about 11\,kpc) at the 25~mag/arcsec$^2$ blue isophote, an axis ratio
close to 2.0, and a position angle of 92\degr\ \citep{Vaucouleurs91}.
\citet{Heisler94}, based on optical imaging of NGC\,1377 as part of a sample
of galaxies with $F_{60\,\mu m} > F_{100\,\mu m}$, mentioned the presence
of a dust lane along the southern part of the minor axis, in an otherwise
featureless morphology. The far-infrared (40--120\,$\mu$m) luminosity of NGC\,1377,
as derived from the IRAS 60 and 100\,$\mu$m fluxes, is $4.3 \times 10^9$\,L$_{\sun}$.
The center of the galaxy contains a compact molecular gas reservoir of
$2 \times 10^8$\,M$_{\sun}$ or more, about ten times more massive than
the amount of gas expected to be associated with the dust seen in emission
(see Section~\ref{model}), if a Galactic gas to dust mass ratio is assumed
\citep{Roussel03}. The molecular gas properties are similar to those characterizing
starbursts: the far-infrared to CO flux ratio is seven times higher than in normal
galaxies and the gas is subthermal.

\begin{figure}[!ht]
\resizebox{12cm}{!}{\plotone{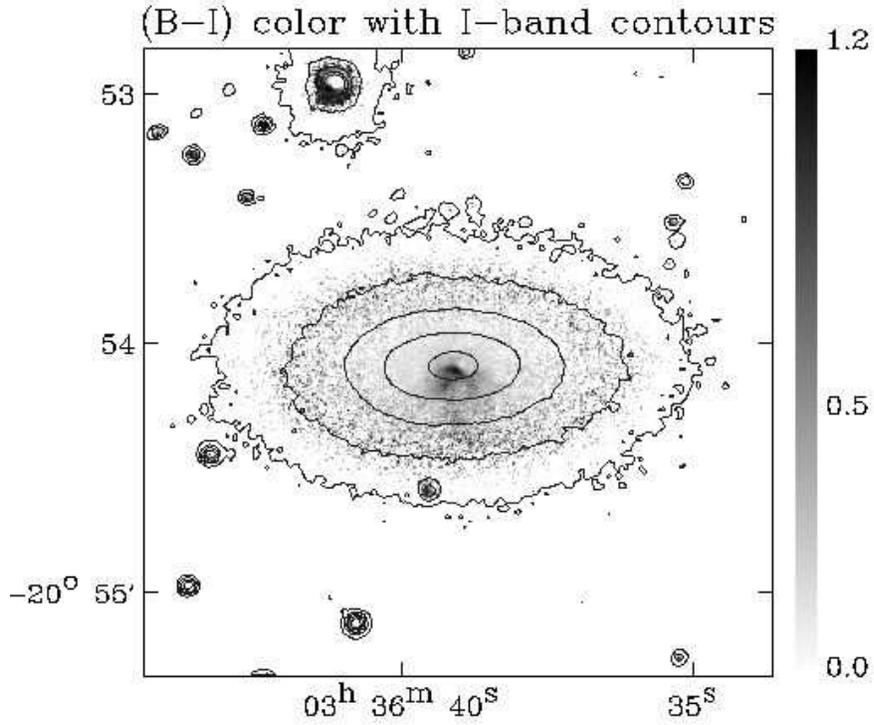}}
\caption{The (B-I) magnitude difference
(with arbitrary zero-points because the maps are not flux-calibrated)
is shown in grey scale, from 0 to 1.2\,mag.
The maximum color in the center of NGC\,1377 is 0.1~mag higher
than displayed here. Isophotes of the I-band image are superposed for the
intensities $\sigma_{\rm I} \times 3^n$, $n = 1$ to 5, where $\sigma_{\rm I}$
is the standard deviation of the sky brightness.
}
\label{fig:color_bi_cont_i}
\end{figure}

\begin{figure}[!ht]
\vspace*{-1.5cm}
\hspace*{-3cm}
\resizebox{21cm}{!}{\plotone{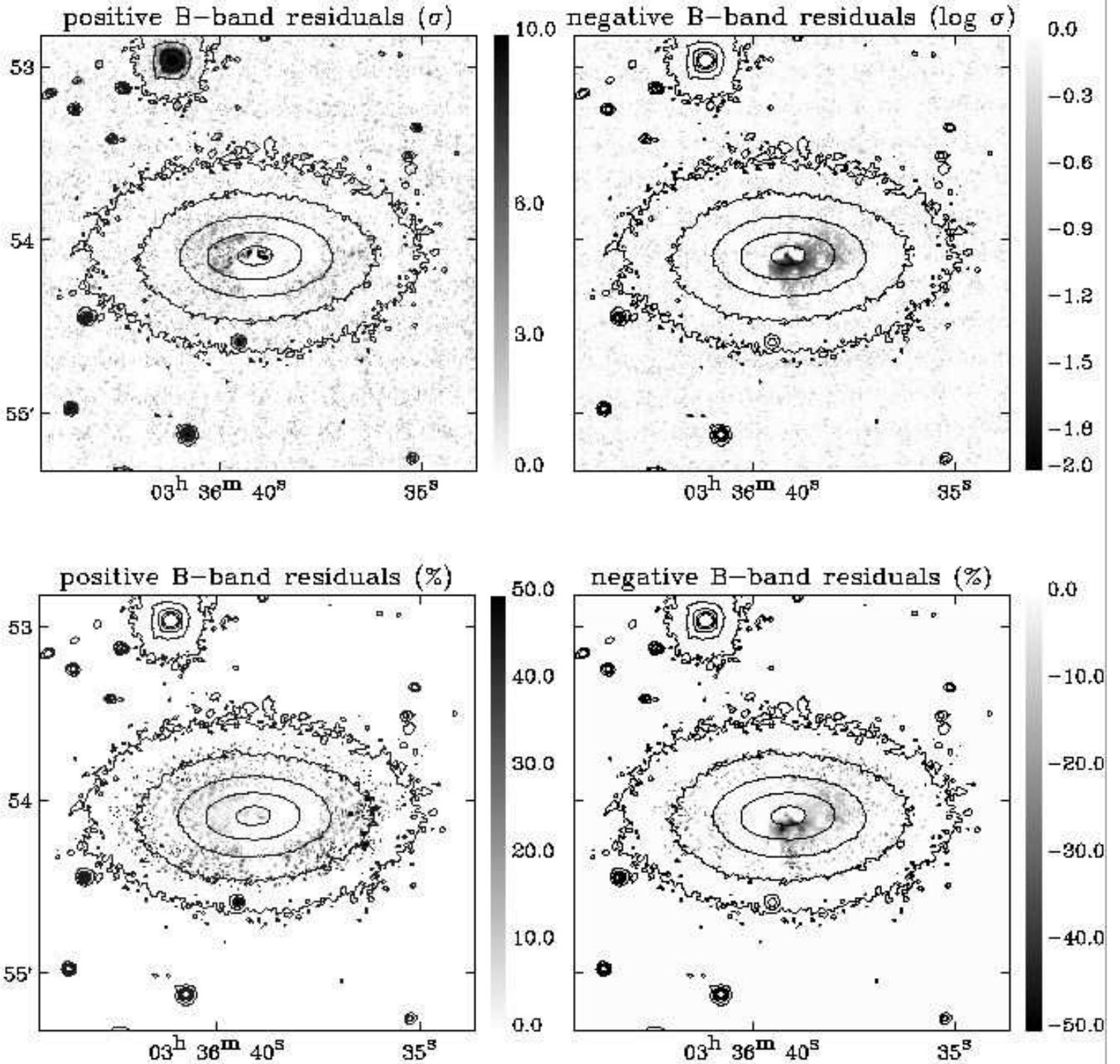}}
\vspace*{-1.5cm}
\caption{
Residuals of the B-band image after subtraction of a smooth photometric
model. To construct this model, radial brightness profiles were derived from small
angular sections along the northern minor axis and along the eastern major axis,
which appear undisturbed in the (B-I) color map. Then, the brightness distribution
at any position angle $\theta$ was interpolated by affecting a weight
cos$^2 \theta$ (sin$^2 \theta$) to the major (minor) axis profile.
{\bf Left:} positive residuals. {\bf Right:} negative residuals.
{\bf Top:} in units of the standard deviation of the sky brightness
(in logarithmic scale for the negative component).
{\bf Bottom:} as a percentage of the local model intensities.
The contours are as in Figure~\ref{fig:color_bi_cont_i}.
}
\label{fig:resi_b}
\end{figure}

Figure~\ref{fig:color_bi_cont_i} shows the (B-I) color map derived from the
optical images obtained for SINGS. The disturbance noted by \citet{Heisler94}
is readily apparent, but seems more complex than a single dust lane. Two main
orthogonal red features are surrounded by more diffuse red filaments. The color
excess between the central feature and the normal light distribution immediately
to the north is approximately 0.9\,mag.
The residuals from the B-band image after subtraction of a smooth symmetric
photometric model are shown in Figure~\ref{fig:resi_b}.
Some faint arcs of
excess light are seen in the center, as well as to the north-east and to
the south-east of the central regions. Negative residuals of large amplitude,
that may be interpreted as regions of excess obscuration and thus excess column
density of cold dust and gas,
are seen in the nucleus and several pockets immediately to the south.
In addition, a straight lane extends further to the south, and diffuse
structures further to the west. These faint residuals are discernible out
to about 2\,kpc in projected distance from the nucleus. The overall morphology
suggests a recent merging event with one or several small bodies,
in which the structure of NGC\,1377 would have been preserved.
As NGC\,1377 is very deficient in atomic hydrogen gas, more robust evidence
cannot be obtained from H{\small I} observations but would require high
spectral resolution and high sensitivity observations of the stellar and
molecular gas kinematics.

\begin{figure}[!ht]
\vspace*{-1cm}
\resizebox{14.5cm}{!}{\plotone{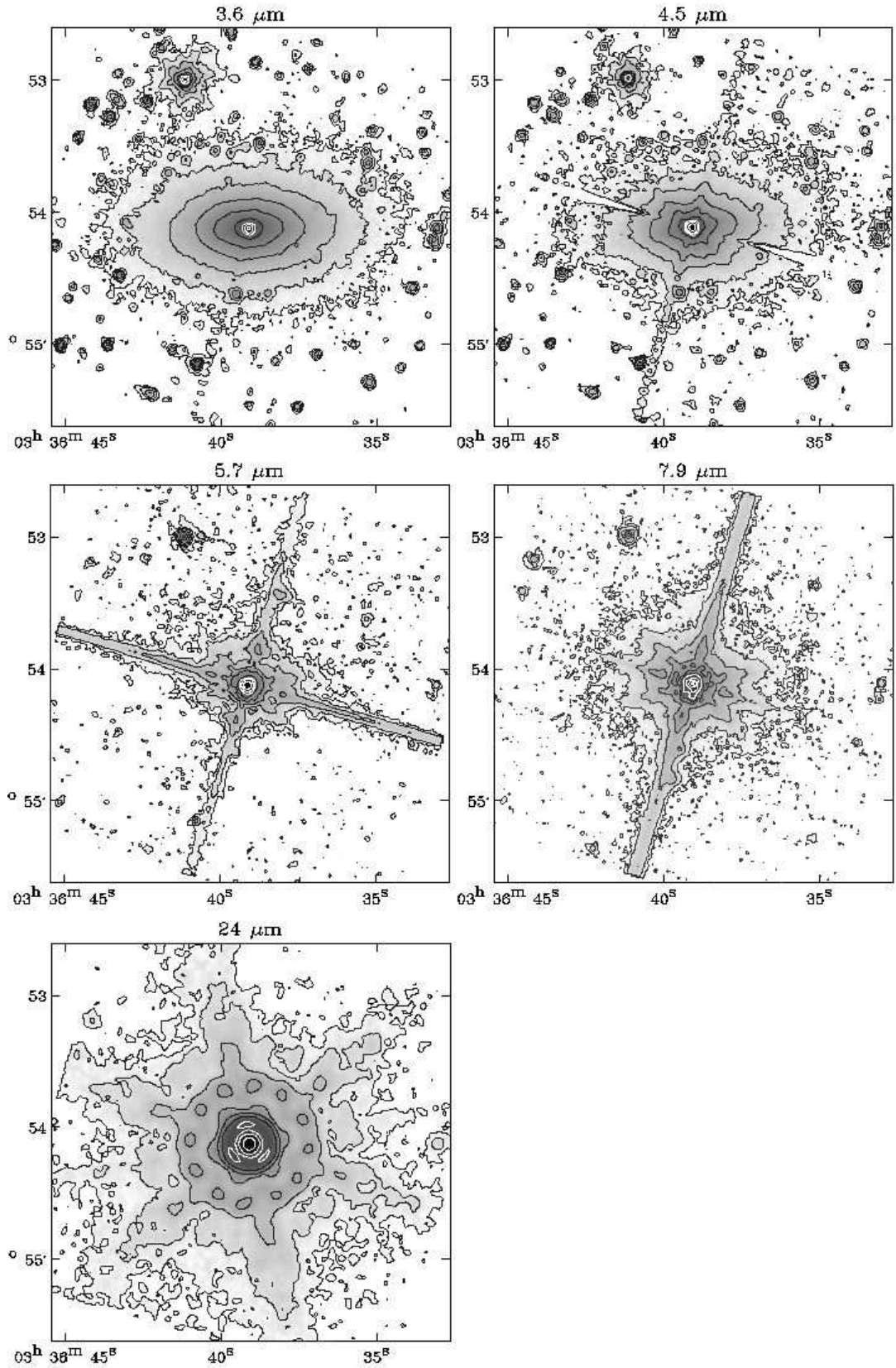}}
\vspace*{-0.3cm}
\caption{Infrared images of NGC\,1377 in the four IRAC bands and in the
24\,$\mu$m MIPS band. The contours are the isophotes $\sigma \times 3^n$,
$n = 1$ to 9 for all bands except 7.9\,$\mu$m, and $n = 1$ to 10 for 7.9\,$\mu$m.
The structure of the images is explained in the text (Sect.~\ref{morph}).
}
\label{fig:imas_ir}
\end{figure}

The infrared images from 3.6 to 24\,$\mu$m are shown in Figure~\ref{fig:imas_ir}.
They are dominated by a very bright nucleus. At 3.6 and 4.5\,$\mu$m,
the underlying smooth emission from the old stellar disk is also visible,
with the same distribution as the I-band light. The reticle-shaped artifact
in the 4.5\,$\mu$m image (with orthogonal positive and negative rows)
is produced by the ``multiplexer bleed'' and ``pull-down'' effects,
respectively, typically associated with bright point sources (IRAC handbook).
At 5.7\,$\mu$m, the stellar disk emission has fallen below the sensitivity
limit, and only the nucleus is detected at longer wavelengths. It has the
appearance of the point response functions of the IRAC and MIPS detectors,
with bright diffraction spikes. The rows that are bright throughout the field
of view of the 5.7 and 7.9\,$\mu$m images exemplify the ``pull-up'' and
``optical banding'' artifacts (IRAC handbook).
Since existing corrections of these artifacts are only cosmetic, we did
not apply them but masked out the affected rows and columns in our analysis.

\begin{figure}[!ht]
\resizebox{12cm}{!}{\plotone{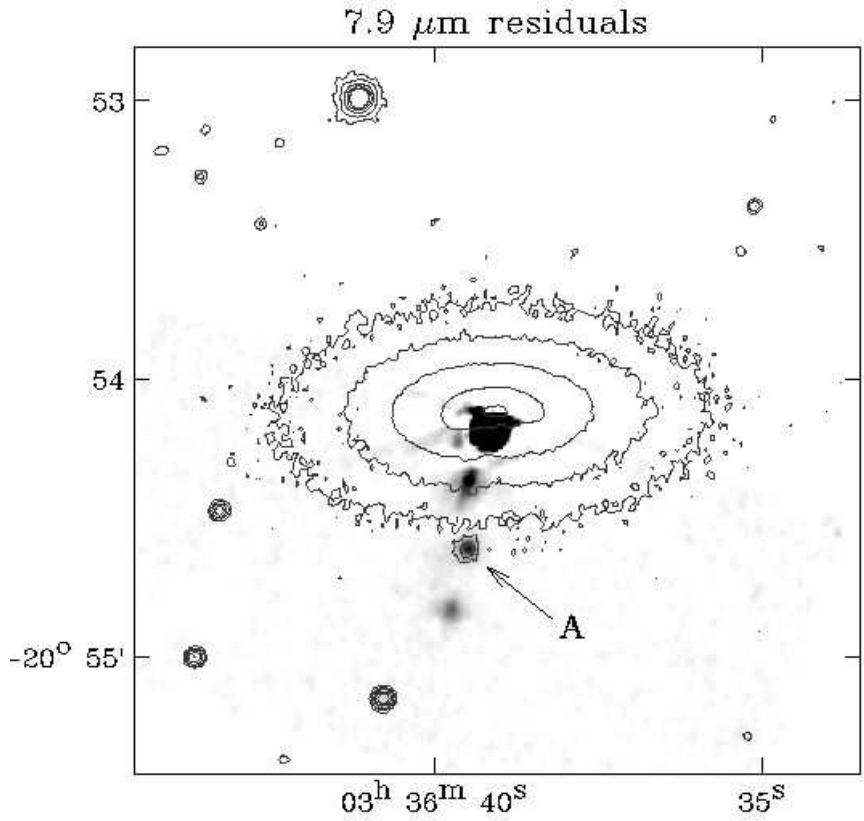}}
\caption{Residuals of the 7.9\,$\mu$m image after subtraction of the mirrored
northern half of the galaxy (see text). An intensity cut of 2\,MJy/sr was applied.
The brightest residuals have an intensity of $\sim 100$\,MJy/sr, and the peak
brightness of the nucleus is about 2000\,MJy/sr. The contours are the B-band
isophotes of intensities $\sigma_{\rm B} \times (1 + 3^n)$, $n = 1$ to 5.
All the residuals are ghosts of the nucleus, except for the source labelled A.
}
\label{fig:resi_8}
\end{figure}

To decompose the nuclear 7.9\,$\mu$m source and determine whether its size
is measurable, point spread functions (PSF) derived from observations of
bright stars are not appropriate, because their cores are more extended than the
observed core of NGC\,1377, sampled at the native IRAC pixel size of 1.22\arcsec.
This is probably because the spectral slope of NGC\,1377 within the 7.9\,$\mu$m
filter bandpass is much steeper than that of stars (the flux density decreasing
sharply from 7.8\,$\mu$m to 9.5\,$\mu$m, because of the deep flanking silicate
absorption), and the PSF width depends on the spectral
slope, increasing as the spectrum reddens.
The model PSFs computed with the tool provided by the Spitzer Science Center,
``S-Tiny Tim'', using the spectrum of NGC\,1377, are also inappropriate for
our purpose, because they depend on pre-launch characteristics and are not accurate
enough.
Within the uncertainties on the PSF, the mid-infrared source in NGC\,1377 is thus
not resolved. Both this constraint and that obtained previously on the size of the
2.12\,$\mu$m molecular hydrogen line emission \citep{Roussel03} imply that the
region of activity is less extended than $\sim 1$\arcsec, i.e. $\sim 100$\,pc.
To look for asymmetries in the 7.9\,$\mu$m emission, we extracted one half
of the map on the left side of the vertical symmetry axis of the diffraction
and ``pull-up'' pattern (the tilted upper half in Figure~\ref{fig:imas_ir}),
and subtracted the mirror image from the other half of the map. The result
is shown in Figure~\ref{fig:resi_8}. The residuals do not include any extended
emission, but primary and secondary ghosts of the nucleus, as well as the real
source labelled A. Given its detection in the optical bands and its approximate
fluxes in the IRAC bands, this source contains both stellar and dust components,
and is likely a distant galaxy.

\section{Optical spectroscopy}
\label{optspec}

\begin{figure}[!ht]
\resizebox{16cm}{!}{\rotatebox{90}{\plotone{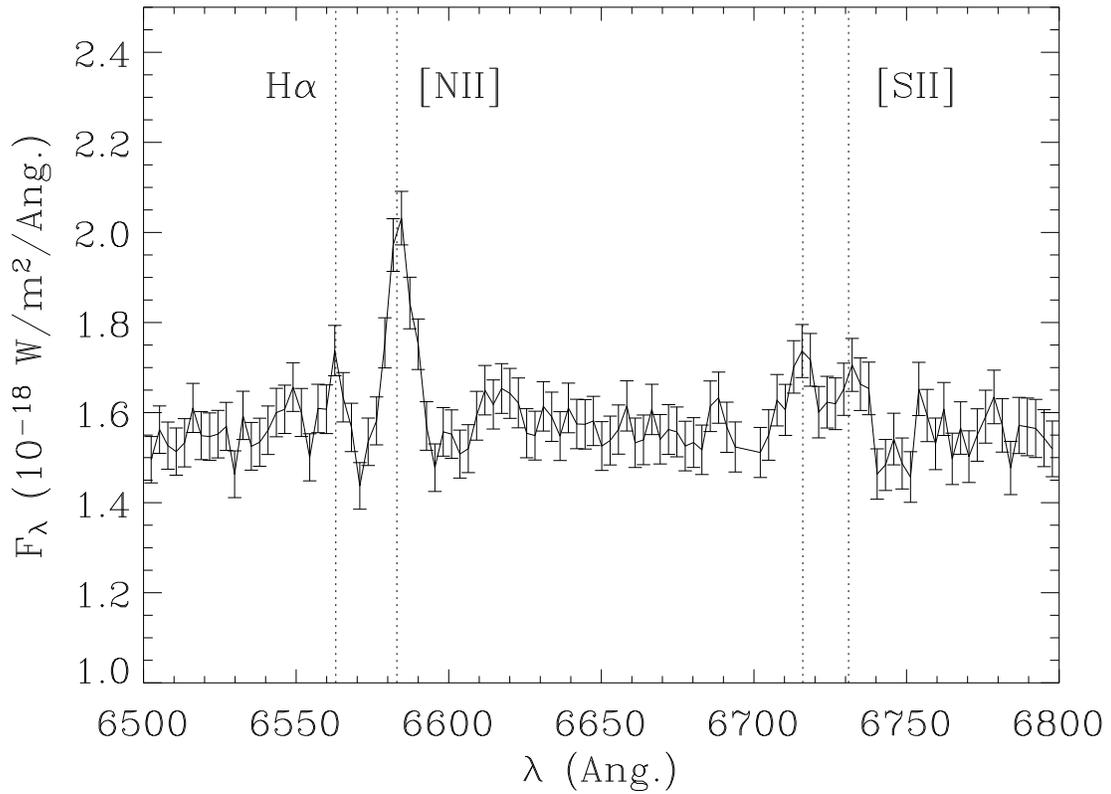}}}
\caption{Rest-frame optical spectrum in the 6500--6800\AA\ region, extracted in a
$2.5\arcsec \times 10\arcsec$ aperture centered on the nucleus. The vertical dotted
lines indicate the wavelengths of the H$\alpha$, [N{\small II}] 6583\AA,
[S{\small II}] 6716\AA\ and [S{\small II}] 6731\AA\ lines.
}
\label{fig:specopt}
\end{figure}

Optical slit spectroscopy shows the [N{\small II}] line at 6583\,\AA\ and the
[S{\small II}] lines at 6716 and 6731\AA\ in emission without any detection of
H$\alpha$ \citep{Kim95}. Our long-slit data confirm this result (Fig.~\ref{fig:specopt}). 
The [O{\small III}], [O{\small II}] and H$\beta$ emission lines are not detected
either. As the optical depth of the infrared source is extremely high
(Sect.~\ref{model}), we infer that [N{\small II}] and [S{\small II}] arise
in the foreground. Even accounting for underlying stellar absorption in the
H$\alpha$ line, the
[N{\small II}](6583\AA)/H$\alpha$
ratio is constrained
to be larger than unity. Since the lines cannot be excited by an active nucleus,
because they do not probe the nuclear regions, they indicate shock excitation
\citep{Baldwin81}. In addition, the low
[S{\small II}](6731\AA)/[S{\small II}](6717\AA)
ratio implies that they
arise in regions of low electronic density \citep{Dopita77}.

Nebular diagnostics of
metallicity are not useable in NGC\,1377. Stellar absorption indices of metallic
lines and H$\beta$, on the other hand, are measureable in the spectrum of NGC\,1377.
But they are degenerate
with respect to age and metallicity, except for galaxies which have
been passive for several gigayears, which appears not to be the case for NGC\,1377:
EW(Mg$_{\rm b}$) = 2.4\AA, EW(Fe$_{5270}$) = 2.1\AA\ and EW(H$\beta$) = 3.3\AA\
in the index definition of \citet{Worthey94}. However, the large quantity of
CO gas and dust (Sect.~\ref{model}), relative to the stellar emission, argues
against NGC\,1377 being a low-metallicity system.

The optical spectrum, covering a large wavelength interval (between 3600 and 6900\AA),
enables us to constrain the star formation history of NGC\,1377. We used the
population synthesis models of \citet{Bruzual03} to fit the entire spectrum
by a discrete sum of instantaneously-formed populations, as described by
\citet{Moustakas06}, assuming solar metallicity and a \citet{Salpeter55} initial
mass function. The estimated stellar extinction is $A_{\rm V} \sim 0.9$~mag in the
screen hypothesis. We find that the light is dominated by populations older than
1\,Gyr ($\sim 60$ to 80\% in the V band), and that the most recent star formation
episode traceable in the optical occurred more than 500\,Myr ago.
This is fully consistent with the radio continuum upper limit indicating that
the cosmic ray population generated by the last major star formation episode
has decayed, implying that this episode occurred more than 100\,Myr ago.

\section{Infrared spectral energy distribution and spectral features}
\label{sed}

\begin{figure}[!ht]
\vspace*{-1.5cm}
\resizebox{14cm}{!}{\rotatebox{90}{\plotone{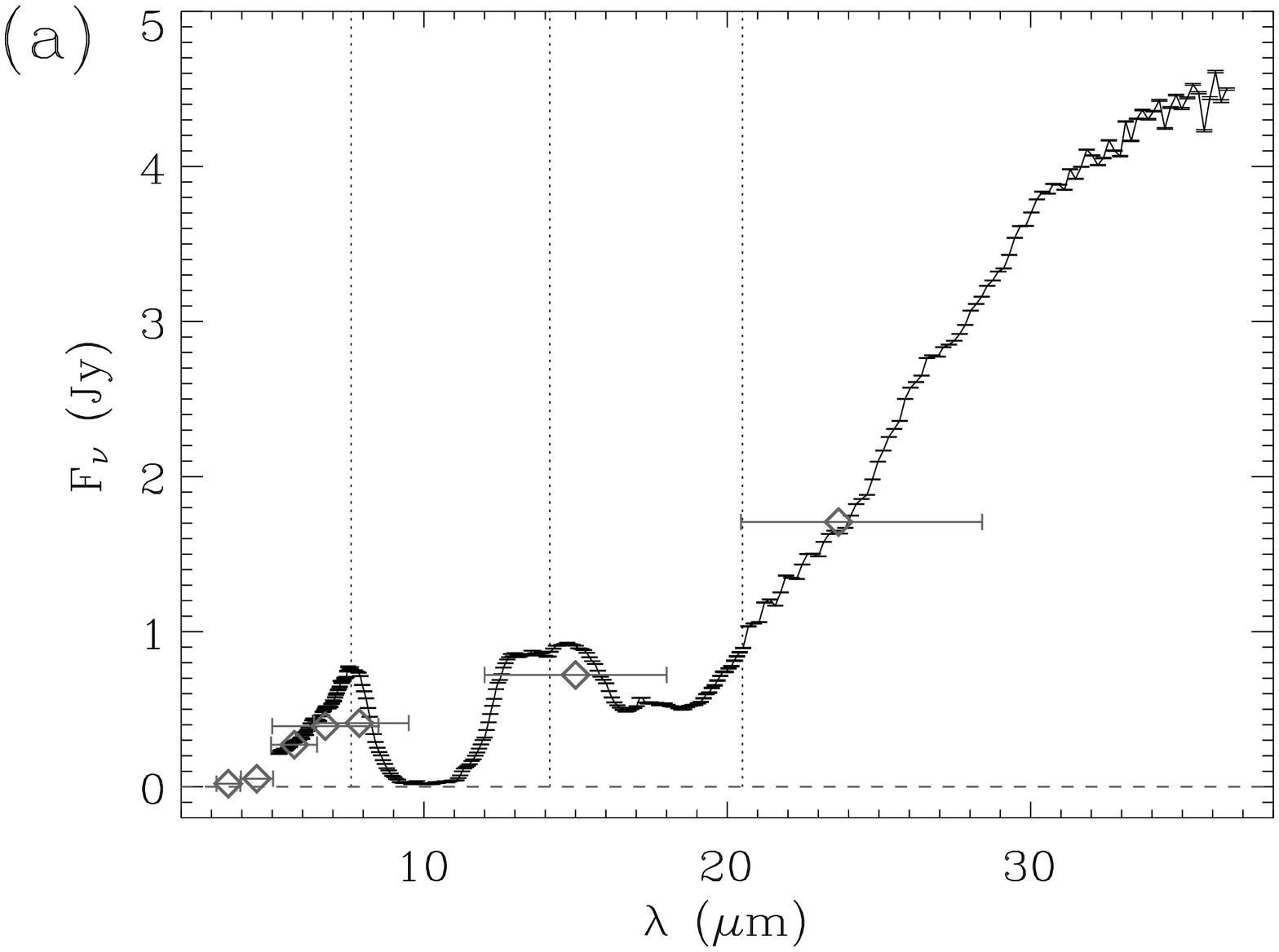}}}
\vspace*{-1cm} \\
\resizebox{14cm}{!}{\rotatebox{90}{\plotone{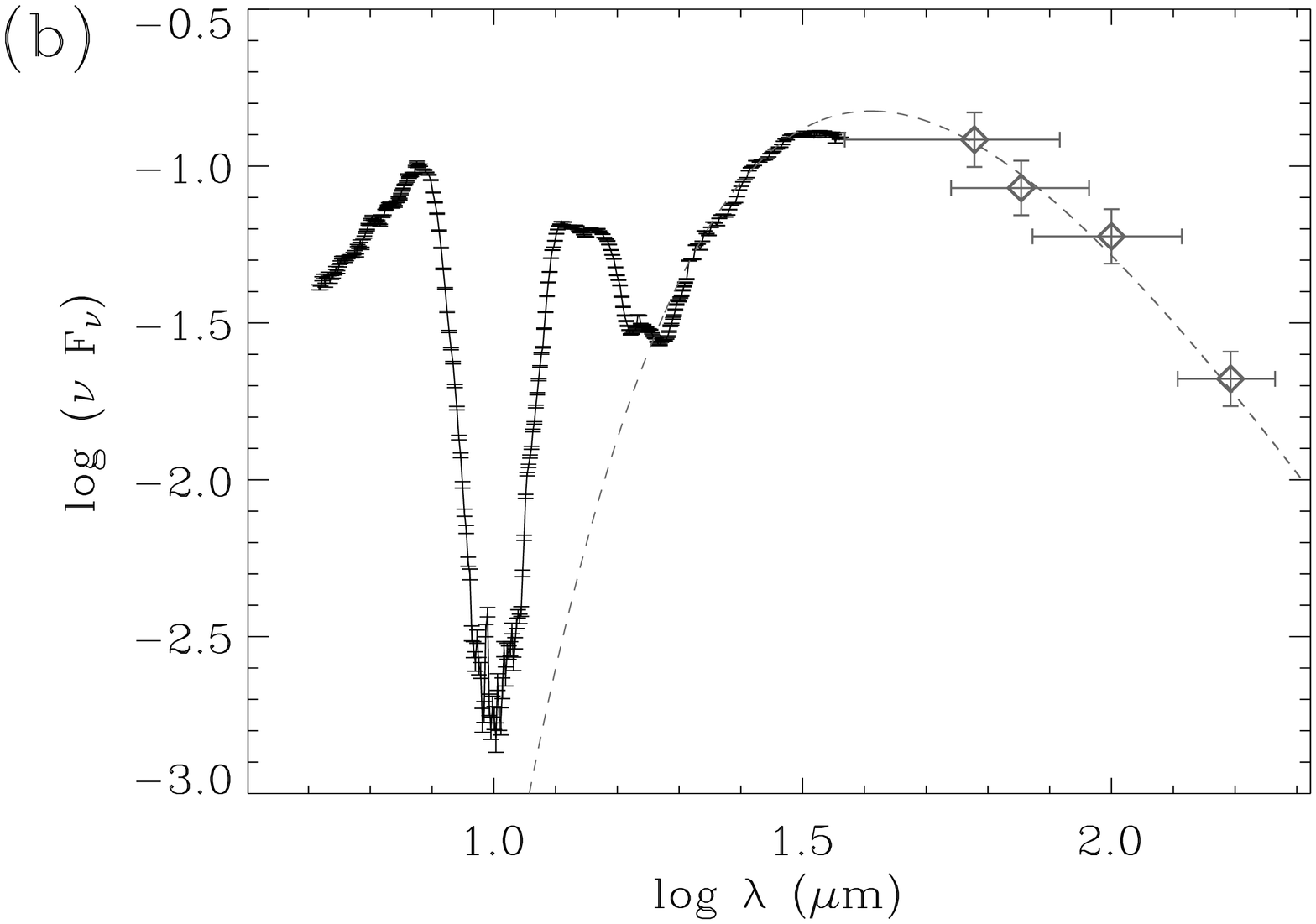}}}
\vspace*{-0.5cm}
\caption{Spectral energy distribution of NGC\,1377 between {\bf (a)} 3 and 40\,$\mu$m,
in flux density units~; {\bf (b)} 5 and 160\,$\mu$m, in power units and logarithmic
scale.
Notice the very deep absorption bands from amorphous silicates at 10 and 18\,$\mu$m.
Broadband imaging photometry from IRAC, ISOCAM (6.75 and 15\,$\mu$m), MIPS
and IRAS (60 and 100\,$\mu$m) is overplotted as grey diamond symbols with horizontal
error bars representing the filter widths. In {\bf (a)}, the spectra from different
orders (SL2, SL1, LL2 and LL1 in order of increasing wavelength) are separated by
vertical dotted lines. In {\bf (b)}, the spectrum from a 90\,K blackbody is
overplotted as a grey dashed line.
}
\label{fig:sed}
\end{figure}

The entire infrared spectral energy distribution, including the low-resolution
spectra, calibrated to reproduce the broadband photometry as explained in
Section~\ref{calib}, is shown in Figure~\ref{fig:sed}. To the far-infrared broadband
fluxes, color corrections between 1.00 and 1.08 were applied. It is immediately
apparent that the infrared source is not only very hot and thus compact
(in agreement with the fact that it is not spatially resolved), but is also
strongly self-absorbed, as shown by the deep absorption bands from amorphous
silicates at 10 and 18\,$\mu$m. The forbidden lines from H{\small II} regions
and photodissociation regions that are bright in star-forming galaxies
([Si{\small II}], [S{\small III}], [Ne{\small II}], [Ne{\small III}]) are absent
in NGC\,1377 (Table~\ref{tab_lines}), except for a very weak [Ne{\small II}]
line at 12.8\,$\mu$m, which will be discussed in Section~\ref{lines}.
The energy peak lies between 30 and 50\,$\mu$m (Fig.~\ref{fig:sed}b), which is
extremely unusual, even for starburst galaxies. An even more extreme spectrum
(with an energy peak around 15-20\,$\mu$m) has been observed in SBS\,0335-052,
a blue compact dwarf galaxy of very low metallicity \citep{Houck04b}.
Since SBS\,0335-052 contains one of the most deeply embedded super star clusters
known, it may be instructive to compare it with NGC\,1377. Their spectra suggest
that the geometry of the source may be similar in both systems, compact and
sharply bounded. There are however important differences. The mid-infrared
emission of SBS\,0335-052 between 5 and 10\,$\mu$m is very faint, indicating that
dust species related to the aromatic band carriers are absent
and that very small grains are depleted with respect to big grains \citep{Plante02}.
On the contrary,
the mid-infrared emission of NGC\,1377 accounts for a large fraction of the total
infrared power and implies that very small grains are abundant (even though aromatic
band carriers are seemingly absent), or alternatively that the radiation field
intensity is high enough to heat big grains to very high temperatures. The optical
depth in the amorphous silicate absorption bands is also much higher in NGC\,1377.

\begin{figure}[!ht]
\vspace*{-3cm}
\hspace*{-1cm}
\resizebox{16cm}{!}{\plotone{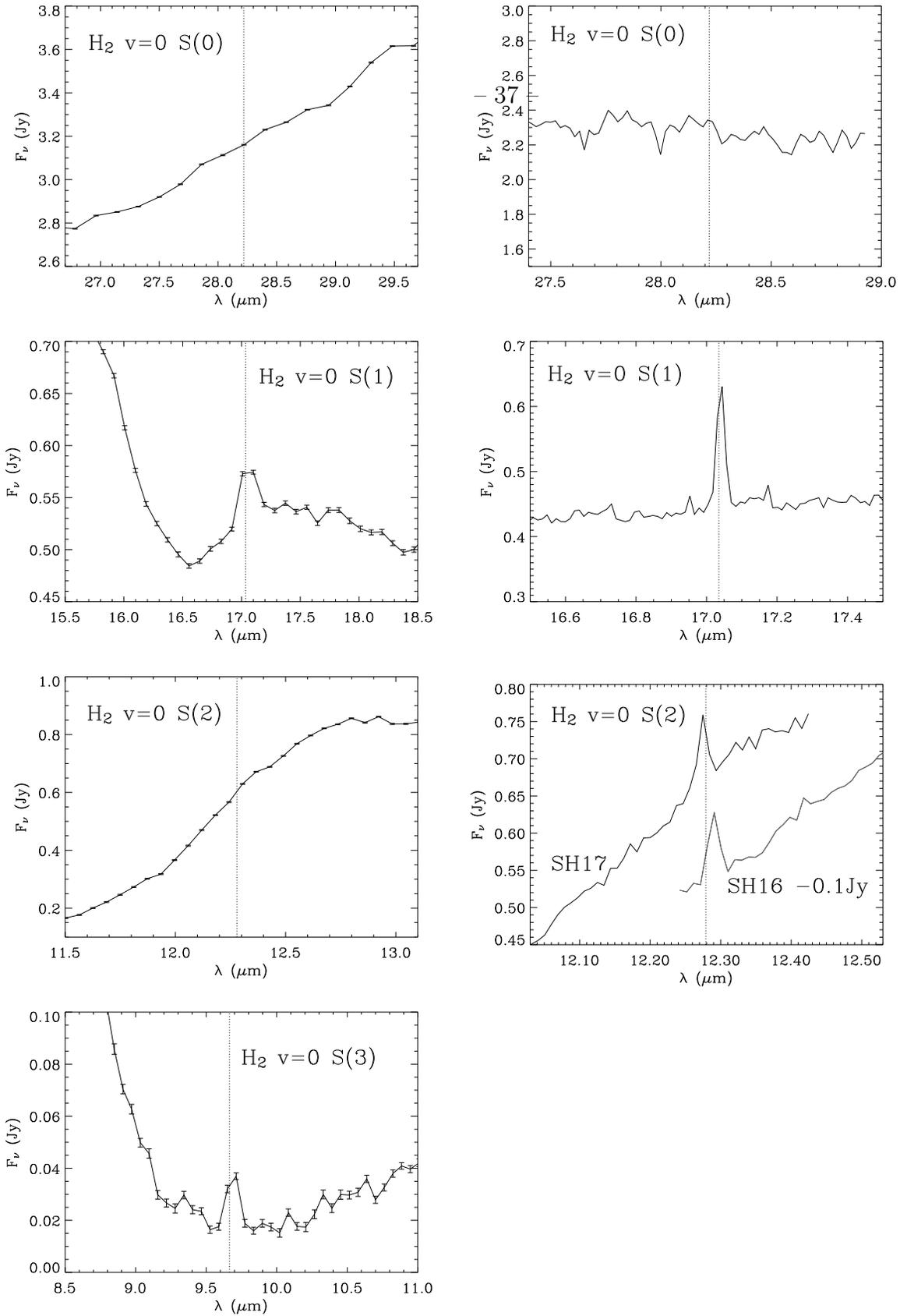}}
\vspace*{-0.3cm}
\caption{H$_2$ rotational lines in the mid-infrared spectrum of NGC\,1377.
{\bf Left:} Rest-frame low-resolution spectra.
{\bf Right:} Rest-frame high-resolution spectra.
The vertical dotted lines indicate the wavelengths of the transitions.
The 12.29\,$\mu$m transition is detected in two different spectral orders
at high resolution~; the two spectra are overplotted, shifted in flux for
clarity.
}
\label{fig:h2}
\end{figure}

\begin{figure}[!ht]
\vspace*{-2cm}
\hspace*{-1.5cm}
\resizebox{18cm}{!}{\plotone{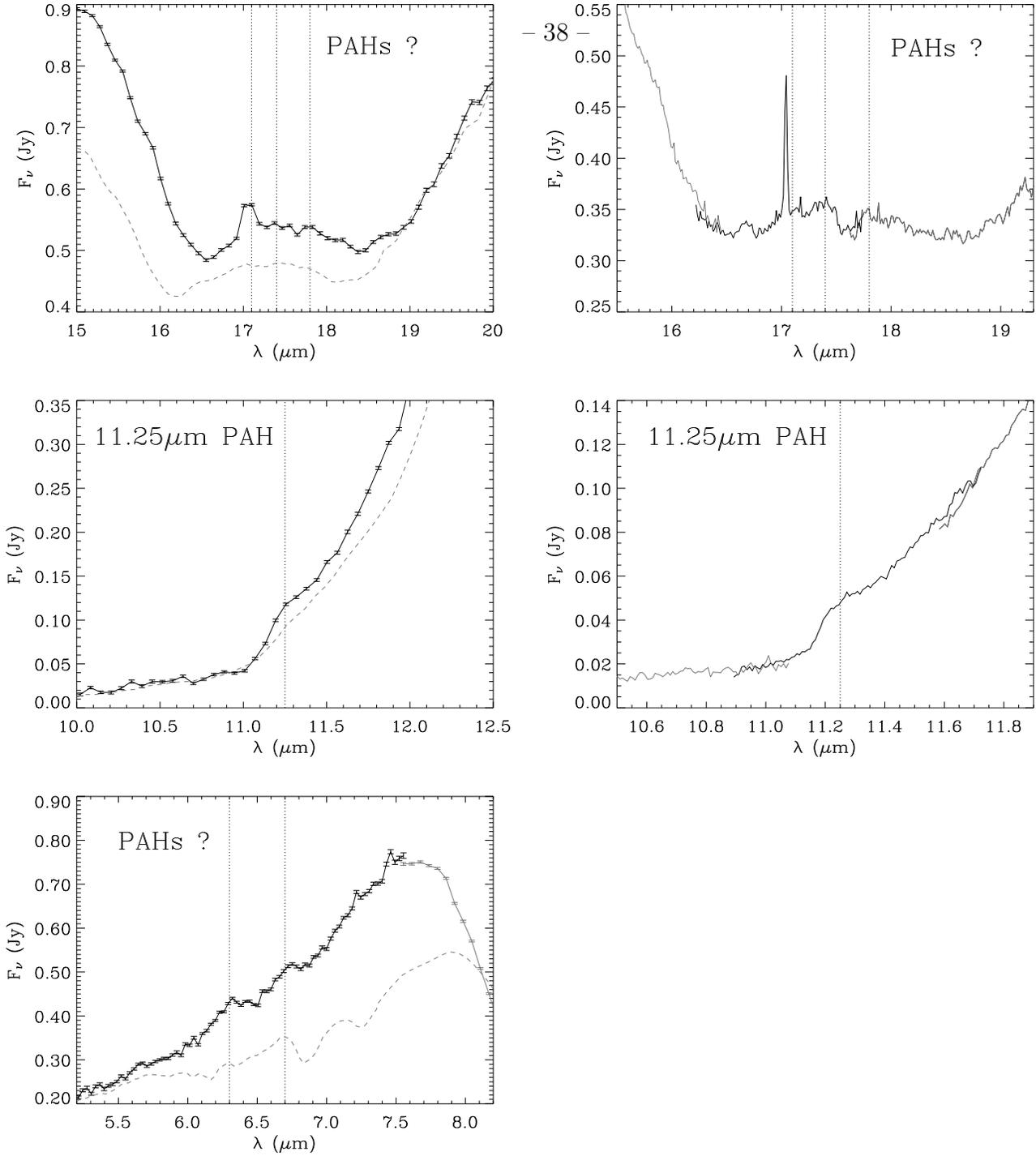}}
\vspace*{-1cm}
\caption{Tentative PAH detections in the mid-infrared spectrum of NGC\,1377.
{\bf Left:} Rest-frame low-resolution spectra.
{\bf Right:} Rest-frame high-resolution spectra.
Different colors are used for different spectral orders.
Spectra in the SH module were extracted in smaller apertures to increase the
signal to noise ratio (in the central $2 \times 2$ pixels for the 11.25\,$\mu$m
feature and $4 \times 4$ pixels for the 17\,$\mu$m complex). For the 17\,$\mu$m
complex, the vertical dotted lines indicate the wavelengths of the
17.1, 17.4 and 17.8\,$\mu$m features. The prominent 17.03 H$_2$ line is not
labelled here and shown separately in Figure~\ref{fig:h2}. In the middle
and bottom spectra, the vertical lines indicate the wavelengths of the 11.3\,$\mu$m,
6.3\,$\mu$m and 6.7\,$\mu$m features.
For each low-resolution spectrum, the spectrum of IRAS\,08572+3915, normalized
to that of NGC\,1377 at 36\,$\mu$m, is overplotted as a grey dashed line (see text).
}
\label{fig:pah}
\end{figure}

No other spectral features are seen, except some molecular hydrogen rotational
lines (Fig.~\ref{fig:h2}), which will be discussed in Section~\ref{gas},
as well as a few broad emission features of very low intensity and weak
absorption features from gas-phase molecules. Superposed on the
deep 10\,$\mu$m silicate absorption band, a very weak 11.3\,$\mu$m emission band
is detected, and likely arises from polycyclic aromatic hydrocarbons
(Fig.~\ref{fig:pah}).
This feature is well matched by the profile of the 11.3\,$\mu$m PAH band
of the Orion bar \citep{Peeters04},
at the spectral resolution
of our data. It may originate from intermediate dust shells, with
optical depths lower than the silicate optical depth, but still higher than
unity (see below).
The 11.3\,$\mu$m aromatic band is always observed in association with other
bands, the most prominent of which lies at 7.7\,$\mu$m. The non-detection of
the 7.7\,$\mu$m band could easily be due to its being hidden in the observed
broad 5--9\,$\mu$m bump, which peaks almost exactly at the central wavelength
of the aromatic band (see Section~\ref{model}). Likewise, a 12.7\,$\mu$m
aromatic band, if present, could be hidden in the broad bump on the red shoulder
of the 10\,$\mu$m silicate band (Fig.~\ref{fig:sed}).

Another broad and faint feature is seen below
the 17.03\,$\mu$m H$_2$ line, at the bottom of the 18\,$\mu$m silicate absorption
band (Figure~\ref{fig:pah}). It could be composed of the same bands at 17.1, 17.4
and 17.8\,$\mu$m that were shown recently to be common in the interstellar medium
of normal galaxies \citep{Werner04, Smith04, Smith06}, and that are thought to be
related to polycyclic aromatic hydrocarbons.
\citet{Kerckhoven00} have extracted a plateau of blended broad features attributed
to the C-C-C bending modes of PAHs in this wavelength range (more exactly between
15 and 20\,$\mu$m), in compact H{\small II} regions as well as young and evolved
stars. They also state that small-size PAHs would have sharper spectral structure
(in particular a band at 16.4\,$\mu$m, unseen in NGC\,1377) than bigger aromatic
compounds. If the 16.5-18.5\,$\mu$m bump observed at the bottom of the 18\,$\mu$m
silicate band in NGC\,1377 is attributable to PAHs, then the lack of substructure
would indicate a predominance of bigger aromatic compounds.

Finally, we tentatively detect two additional broad features at 6.3 and 6.7\,$\mu$m
(Fig.~\ref{fig:pah}). The dip longward of 6.8\,$\mu$m could be caused by an
absorption from C-H bending modes of hydrocarbons at 6.85\,$\mu$m, seen in
deeply obscured infrared-luminous galaxies by \citet{Spoon01} and \citet{Spoon04}.
For comparison, Fig.~\ref{fig:pah} shows the spectrum of one of these galaxies,
IRAS\,08572+3915 \citep{Spoon06a}, where the absorption bands from hydrocarbons
at 6.85\,$\mu$m and 7.25\,$\mu$m are clearly seen.
However, there is no known species being able to cause the dip at 6.5\,$\mu$m.
The 6.3\,$\mu$m band has been observed in at least two post-AGB stars by
\citet{Peeters02} and has also been invoked by \citet{Sturm00} to explain the
excess emission redward of the 6.2\,$\mu$m aromatic band of starburst galaxies.
It has so far not been observed alone (without the 6.2\,$\mu$m band) in galaxies.
According to \citet{Peeters02}, the class of PAHs with a 6.3\,$\mu$m peak wavelength
may have a more pristine composition than the 6.2\,$\mu$m PAHs, and may not yet
have been exposed to hard radiation. It can be expected from the compact starburst
nature of NGC\,1377 that PAHs have all been
destroyed in the zone of activity and survive only in outer shells where they
would not yet have felt the influence of the starburst~; it would however be
difficult to understand how they could have avoided some kind of processing.
The as yet unidentified 6.7\,$\mu$m band has been observed in NGC\,7023
\citep{Werner04} and in spiral galaxies \citep{Smith06}.

If we assume that the intrinsic relative intensities of the 6.3, 6.7, 17--18 and
11.3\,$\mu$m bands are similar to what is seen in normal galaxies, then they are
not consistent with zero extinction, but would be consistent with an optical
depth of the order of half the optical depth affecting the continuum (which is
about 8.5 at 11.3\,$\mu$m, as constrained in Sect.~\ref{model}). Given the
utter uncertainty on the relative geometry of the different dust phases, and
the fact that the relative intensities of the aromatic bands are intrinsically
variable in various interstellar sources and in galaxies
\citep{Peeters02, Smith06}, the identification of the reported features with PAHs
is therefore plausible.

\begin{figure}[!ht]
\resizebox{14cm}{!}{\rotatebox{90}{\plotone{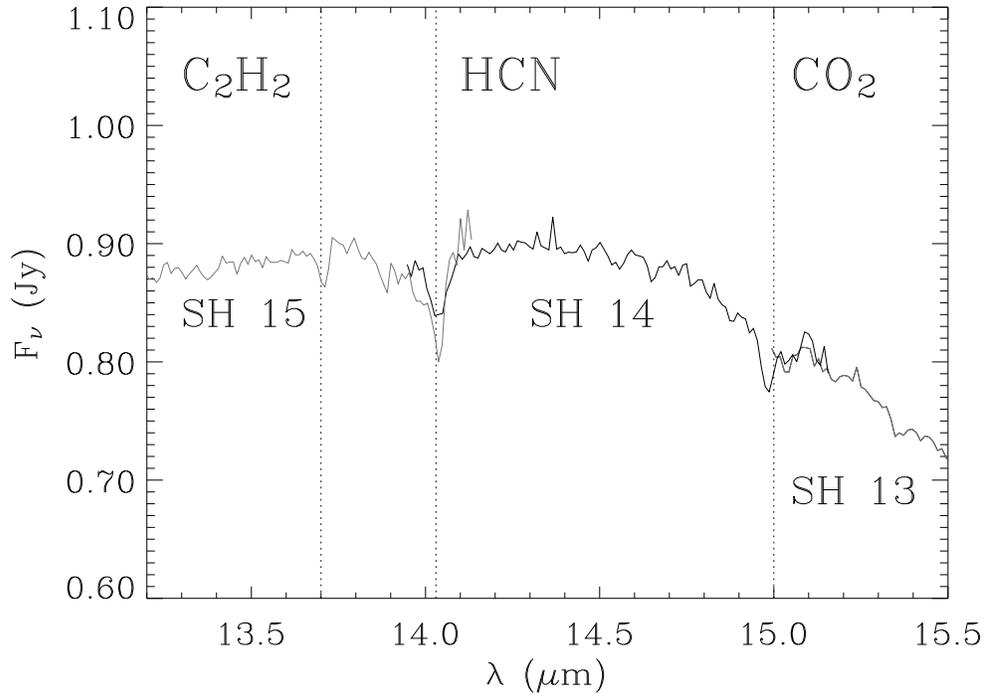}}}
\resizebox{14cm}{!}{\rotatebox{90}{\plotone{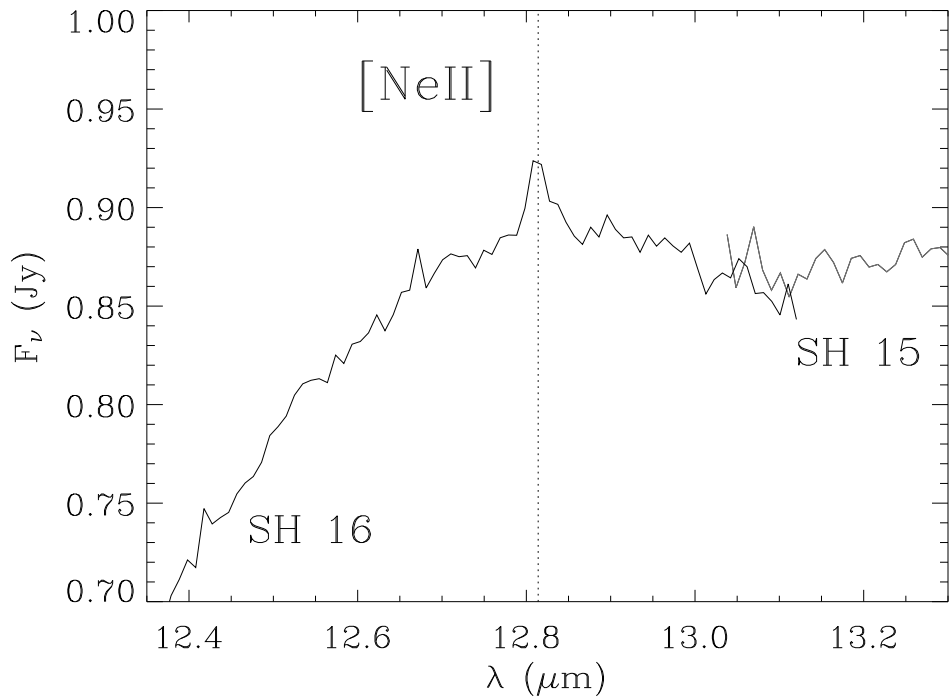}}}
\caption{Rest-frame high-resolution spectra of faint molecular absorption features
and of the [Ne{\small II}] line. Different colors are used for different spectral
orders. The vertical dotted lines indicate the wavelength of the features.
}
\label{fig:sh_features}
\end{figure}

Very faint absorption features are seen at high resolution between the two
amorphous silicate bands: HCN at 14.03\,$\mu$m, which lies at the end
of two spectral orders but is confirmed in both, possibly C$_2$H$_2$
at 13.7\,$\mu$m, and CO$_2$ at 15.0\,$\mu$m, which lies at the end
of a single spectral order and is outside the wavelength range of the adjacent
order (Fig.~\ref{fig:sh_features}). These absorption bands, tracers of hot
dense gas, have been observed in deeply obscured infrared-luminous galaxies
\citep{Spoon06b}. Contrary to what is seen in some of the latter objects and
in Galactic center sources \citep{Chiar00}, we do not detect any absorption
features in the 5-8\,$\mu$m range, i.e. from water ice at 6\,$\mu$m and from
hydrocarbons at 6.85 and 7.25\,$\mu$m.

\section{Geometry and density}
\label{model}

To attempt to constrain the geometry of the infrared source in NGC\,1377,
following the same approach as \citet{Plante02} for SBS\,0335-052, we used the
{\small DUSTY} radiative transfer model \citep{Nenkova00}, assuming a unique,
spherically-symmetric source. By construction, aromatic compounds are not included
in the model, and impulsive heating\footnote{also referred to as transient,
stochastic, or single-photon heating} of very small grains is not taken into account.
This shortcoming is partially
mitigated by the fact that the dust in NGC\,1377 is heated
to very high temperatures, and thus closer to thermal equilibrium than in normal
starbursts, but the results should be interpreted with caution. However, we note
that the effect of neglecting impulsive heating should produce residuals smoothly
distributed\ in wavelength, not narrow features (see below).
Another important limitation is that the same grain size distribution has to be
postulated for all dust species, despite the fact that all the families of dust
models of the local interstellar medium need very different size distributions
for silicates and carbon-based dust \citep{Zubko04}. Since no unique set
of geometric parameters will provide a fit to the data, and extensive exploration
of the parameter space would be prohibitively long for such a high optical depth
as seen in NGC\,1377, we restrict the analysis to a qualitative assessment of
the density. For the silicate grains, we find that, among the choices available
within {\small DUSTY}, the optical properties of \citet{Ossenkopff92} allow
the best reproduction of the shape of the silicate absorption bands. For the
carbonaceous grains, graphite, with the optical properties of \citet{Draine84},
allows a better fit to the spectrum of NGC\,1377 than amorphous carbon.

\begin{figure}[!ht]
\hspace*{-1.5cm}
\resizebox{16cm}{!}{\rotatebox{90}{\plotone{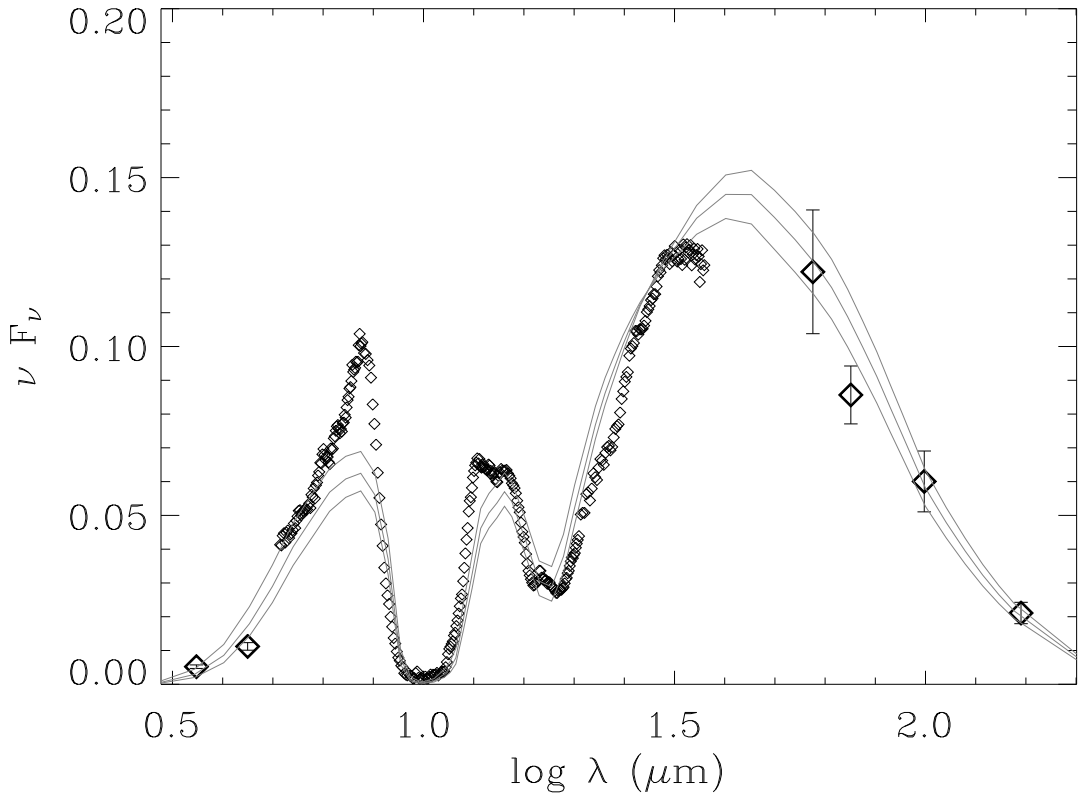}}}
\caption{The fit to the infrared spectral energy distribution of NGC\,1377
by the DUSTY model, with the parameters discussed in the text, is overplotted
in grey, for $\tau_{\rm V} = 75$, 85 and 95. Broadband fluxes are shown as
big diamonds with their error bars. Color corrections between 1.00 and 1.08
have been applied to the far-infrared fluxes (60, 71, 100 and 156\,$\mu$m)
and corrections of 0.92 and 0.96 to the near-infrared fluxes (3.6 and 4.5\,$\mu$m).
}
\label{fig:fit_dusty}
\end{figure}

Figure~\ref{fig:fit_dusty} shows fits reproducing the available data from 3.6
to 156\,$\mu$m, except in a few wavelength intervals that will be discussed
separately below. The models contain at least 70\% of amorphous silicates
by number of grains, which corresponds to about 80\% of silicates by mass,
using the mass densities provided by \citet{Weingartner01} and \citet{Laor93}.
This mass abundance is a bit larger than required by the relevant class
of dust models of \citet{Zubko04} ($\sim 66$\%), as well as by the dust model of
\citet{Li01} ($\sim 73$\%). Such a large fraction of silicates is made necessary
in NGC\,1377
to account for the great depth of the 10\,$\mu$m and 18\,$\mu$m features.
It can be speculated that this reflects composition changes effected by the
intense radiation in the starburst region,
silicates possibly being more resilient
than carbonaceous grains in an oxygen-rich medium.
The best fits are also obtained for a grain size distribution enhanced in small
grains, with a power-law index $-5$ instead of $-3.5$ (for the standard
\citet{Mathis77} distribution) and a lower cutoff of the order of 10\,\AA.
In particular, for a normal size distribution, the opacity would have to be much
higher in order to reproduce correctly the 10\,$\mu$m and 18\,$\mu$m absorption bands.
This may be an artifact of the model, compensating for the absence of impulsive
heating.

The model implies an opacity of at least 75 in the V band, more than 20
at the center of the 10\,$\mu$m silicate band, and about 10 in the 18\,$\mu$m
silicate band. The radial density profile, that was parameterized as a broken
power-law, is very steep, with an index of $-2$ in the center and $-1.5$ outside.
The density profile in the outer parts is ill-constrained, and does not allow us
to derive a total size. For a bolometric luminosity of
$1.1 \times 10^{10}$\,L$_{\sun}$, derived from the data themselves
by integration over the whole infrared spectral energy distribution,
and a temperature of 2000\,K at the inner surface of the dust shell, the radius
of this inner layer is constrained to be 0.08\,pc, very similar to what
\citet{Plante02} found for SBS\,0335-052, although the total infrared luminosity
of NGC\,1377 is about three times higher (the inner dust temperature of the model
of SBS\,0335-052 is only 700\,K).
The dust mass derived from the 60\,$\mu$m and 100\,$\mu$m fluxes, using the Galactic
emissivity as formulated by \citet{Bianchi}, is $2.7 \times 10^5$\,M$_{\sun}$.
It can alternatively be derived from the opacity obtained with the DUSTY model:
\begin{eqnarray*}
\tau_{\rm V} = \sum_{\rm i} (f_{\rm i}~ \kappa_{\rm V\,i}~ \rho_{\rm i}) \int_a (4/3 \pi a^3~ n(a)\,da) \int_r (N(r)\,dr) \\
{\rm and}~~ M_{\rm dust} = \sum_{\rm i} (f_{\rm i}~ \rho_{\rm i}) \int_a (4/3 \pi a^3~ n(a)\,da) \int_r (N(r)~ 4 \pi r^2\,dr),
\end{eqnarray*}
where the index $i$ designates each dust species (here silicates and graphite),
$f_{\rm i}$ is their fraction by number, $\kappa_{\rm V\,i}$ is their mass absorption
coefficient in the V band, $\rho_{\rm i}$ is the solid density of the grains,
$n(a)\,da$ the grain size distribution and $N(r)\,dr$ the radial density distribution
adopted in the model. It follows that
\begin{eqnarray*}
M_{\rm dust} = 4 \pi~ \tau_{\rm V}~ \frac{\sum_{\rm i} (f_{\rm i}~ \rho_{\rm i})}{\sum_{\rm i} (f_{\rm i}~ \kappa_{\rm V\,i}~ \rho_{\rm i})}~ \frac{\int_r (N(r)~ r^2\,dr)}{\int_r (N(r)\,dr)}.
\end{eqnarray*}
We used $\kappa_{\rm V} = 0.3$\,m$^2$\,g$^{-1}$ for silicates and 5\,m$^2$\,g$^{-1}$
for graphite \citep{Draine84}~; and $\rho = 3.5$\,g\,cm$^{-3}$ for silicates and
2.24\,g\,cm$^{-3}$ for graphite \citep{Weingartner01, Laor93}.
Varying the V-band optical depth between 75 and 95 (Fig.~\ref{fig:fit_dusty}),
we obtain a dust mass in the range (2.6--3.3$) \times 10^5$\,M$_{\sun}$, in
good agreement with the determination from the far-infrared fluxes.
Assuming a gas to dust mass ratio of 100 \citep{Sodroski94}, we obtain a central
gas density of about $3 \times 10^4$\,M$_{\sun}$\,pc$^{-3}$, or equivalently a
hydrogen nucleus density of $10^4$\,cm$^{-3}$.

It should be noted here that the observations up to 156\,$\mu$m are not sensitive
to the potential presence of very cold dust (not illuminated by the starburst).
Our dust mass estimates are therefore lower limits, and we suspect the total dust
mass to be about ten times higher, since the mass of cold molecular hydrogen
is ten times higher than the mass of gas expected to be associated with the dust
seen in emission.

\begin{figure}[!ht]
\hspace*{-2cm}
\resizebox{20cm}{!}{\rotatebox{90}{\plotone{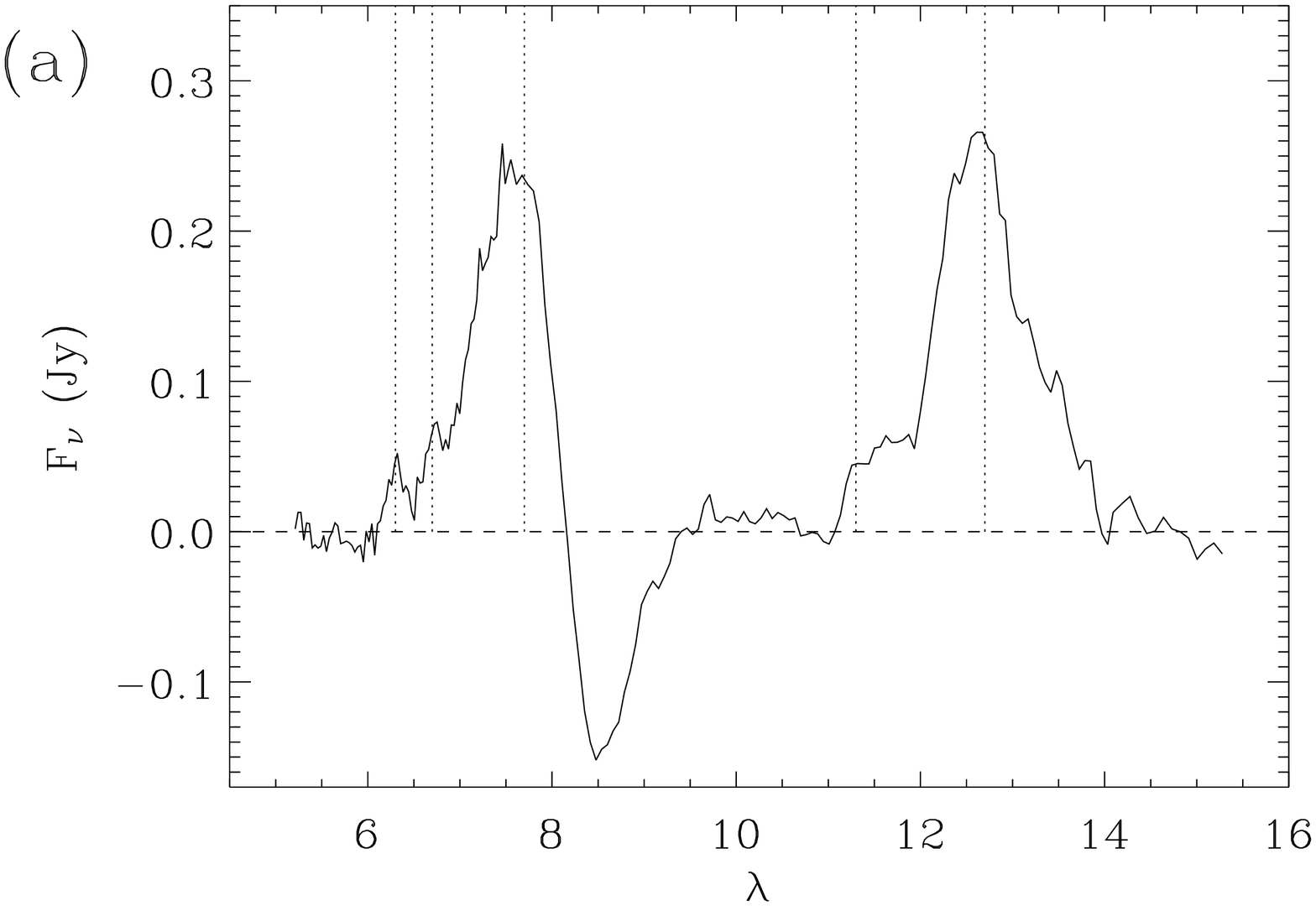}}\rotatebox{90}{\plotone{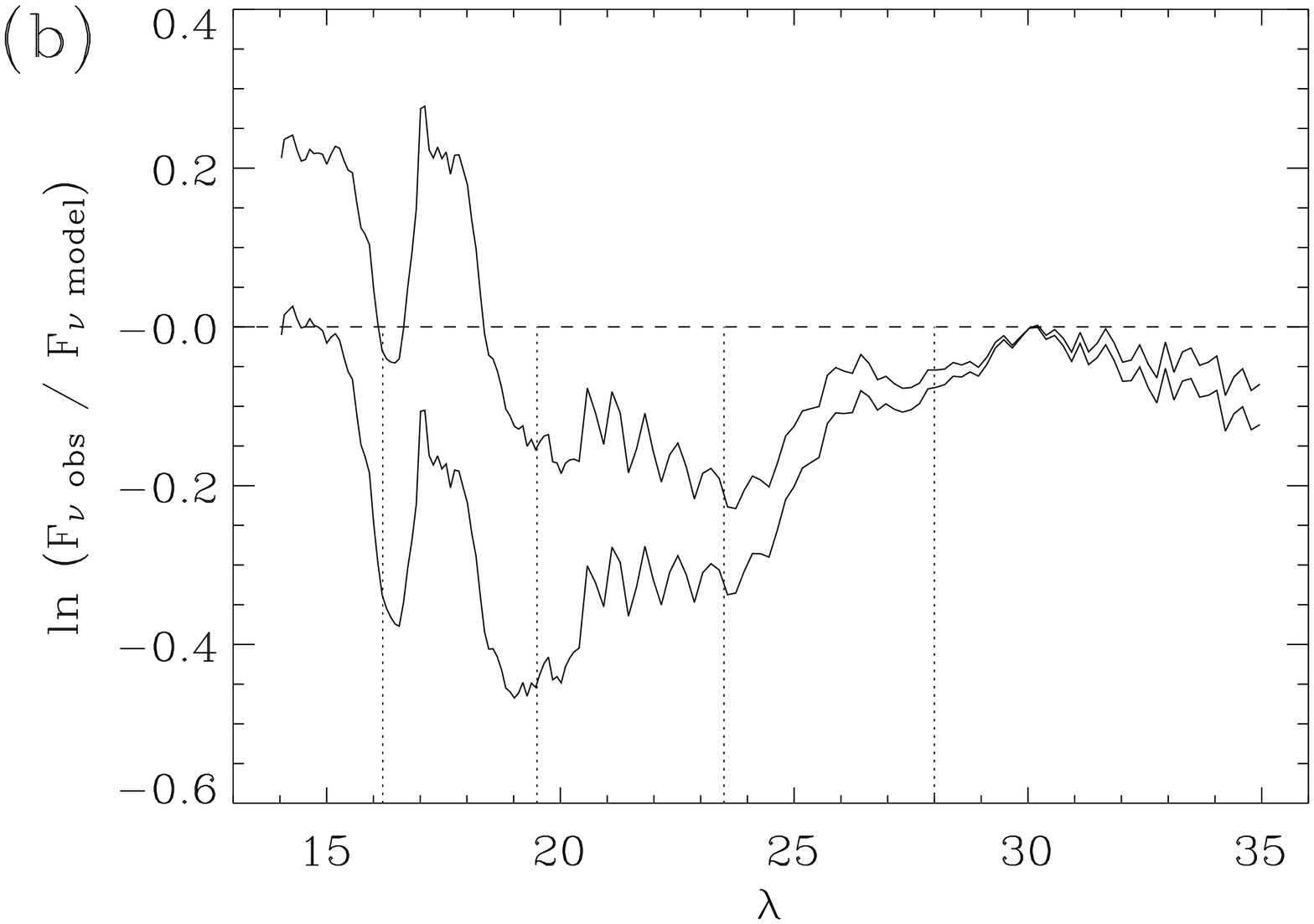}}}
\caption{Residuals after subtraction or normalization by the DUSTY model.
{\bf (a)} Minimum residual emission in the 5--15\,$\mu$m range (for $\tau_{\rm V} = 75$).
The peak wavelengths of the aromatic bands discussed in the text are shown by
vertical dotted lines (at 6.3, 6.7, 7.7, 11.3 and 12.7\,$\mu$m).
We stress that due to modelling limitations, the 7.7 and 12.7\,$\mu$m bands
are very uncertain (see text).
{\bf (b)} Residuals in the 15--35\,$\mu$m range shown for $\tau_{\rm V} = 75$
(lower curve) and $\tau_{\rm V} = 95$ (upper curve), in optical depth scale.
The vertical lines indicate the positions of crystalline silicate features
discussed by \citet{Spoon06a} (at 16, 19.5, 23.5 and 28\,$\mu$m).
The effect of fringing is seen in observations with the LL1 module, particularly
between 20.5 and 24\,$\mu$m, and beyond 30\,$\mu$m.
}
\label{fig:fit_residus}
\end{figure}

The residuals, after subtraction of the model (Fig.~\ref{fig:fit_residus}),
can tentatively be identified with emission bands from aromatic compounds
and absorption bands from crystalline silicates.
If the identification with aromatic band carriers is correct, their chemical
composition is likely different from those dominating the mid-infrared spectrum
of normal star-forming galaxies, as indicated by the absence of the 6.2\,$\mu$m
band near the 6.3\,$\mu$m band (Section~\ref{sed}). In Figure~\ref{fig:fit_residus}a,
prominent features are seen at peak rest wavelengths of 7.7 and 12.7\,$\mu$m.
Note that the absorption profile of silicates is not perfectly reproduced~;
it might be possible to explain the excess emission by a combination of impulsive
heating (neglected in the model) and modified optical properties of silicates.

Absorption bands from crystalline silicates have been found in some
infrared-luminous galaxies by \citet{Spoon06a}. In Figure~\ref{fig:fit_residus}b,
features are seen at rest wavelengths of 23.5 and 28\,$\mu$m. Other known features
from forsterite exist at 16 and 19.5\,$\mu$m, but the quantification of their
depth in NGC\,1377 is made hazardous by the possible contamination by aromatic
bands in the 17--18\,$\mu$m range (Fig.~\ref{fig:pah}). The 16\,$\mu$m band,
in particular, would have to be shifted to longer wavelengths in comparison with
IRAS\,08572+3915. Another feature exists at $\approx 11$\,$\mu$m, but there again
it is confused by the 11.3\,$\mu$m PAH. Crystalline forsterite also produces strong
bands at about 34 and 70\,$\mu$m, among others \citep{Koike93, Molster02}.
The dip of the spectrum of NGC\,1377 at the long-wavelength end of the LL1
module, between 30 and 35\,$\mu$m (Fig.~\ref{fig:sed} and \ref{fig:fit_residus})
may be attributed to absorption by crystalline silicates. Likewise, they may
contribute, at least partially, to the weakness of the 71\,$\mu$m broadband
flux relative to the 60, 100 and 156\,$\mu$m fluxes.

\section{Forbidden lines}
\label{lines}

Mid-infrared forbidden lines are important tracers of H{\small II} and
photodissociation regions, and of the physical conditions therein.
Figure~\ref{fig:sh_features} shows the detection of a very weak [Ne{\small II}]
line. [Ne{\small II}] emission without detection of any other forbidden line
in the mid-infrared range is also observed in some galaxies of the sample of
\citet{Spoon06a}. We attempt here to extract, from the detection of [Ne{\small II}]
and the absence of other lines, constraints on the recombination rate
and electronic density. The low signal to noise ratio of [Ne{\small II}]
precludes measuring the line width accurately, but it is well matched by
the spectral resolution (corresponding to a full width at half maximum of
500\,km\,s$^{-1}$ for an unresolved line).

The estimated flux (Table~\ref{tab_lines})
can be translated into a recombination rate. We have
\begin{eqnarray*}
F_{\rm [NeII]} \times 4 \pi D^2 = {\rm h} \nu~ V~ C~ N({\rm Ne}^+)~ N({\rm e}^-)~ (T_{\rm e} / 1\,{\rm K})^{-0.5}~ \Omega(T_{\rm e})~ g_{\rm l}^{-1}~ exp(-{\rm h} \nu~ /~ (k T_{\rm e}))
\end{eqnarray*}
and $N_{\rm r} = \alpha(T_{\rm e})~ V~ N({\rm H}^+)~ N({\rm e}^-)$, where $D$ is
the distance of NGC\,1377, $V$ is the volume of the H{\small II} region,
$C = 8.6287 \times 10^{-6}$\,cm$^3$\,s$^{-1}$,
$N$ designates a volume density, $\Omega(T_{\rm e})$ is the collision strength,
$g_{\rm l}$ is the statistical weight of the ground level (4 for [Ne{\small II}]
and 5 for [Ne{\small III}]), and $\alpha(T_{\rm e})$, which depends only weakly on
the electronic density, is the hydrogen recombination coefficient. We assume that
$N({\rm Ne}) / N({\rm H}) = [N({\rm Ne}^+) + N({\rm Ne}^{2+})] / N({\rm H}^+)$
and that the neon abundance is equal to the solar value, $6.9 \times 10^{-5}$
\citep{Asplund04}. The collision strengths of the [Ne{\small II}] and
[Ne{\small III}] transitions are tabulated respectively by \citet{Saraph94} and
\citet{Butler94}, and the hydrogen recombination coefficient by \citet{Storey95}.
Accounting for the upper limit on the [Ne{\small III}] line flux and assuming
that the electronic temperature is between $5 \times 10^3$\,K and $10^4$\,K,
we obtain a recombination rate (uncorrected for the extinction in the [Ne{\small II}]
and [Ne{\small III}] lines)
$N_{\rm r} = (5.7 \pm 1.6) \times 10^{51}$\,s$^{-1}$.

The upper limit on the thermal emission at 3\,cm \citep{Roussel03} independently
provides an upper limit
on the recombination rate:
$N_{\rm r} < (T_{\rm e} / 10^4\,{\rm K})^{-0.45} \times 5.4 \times 10^{52}$\,s$^{-1}$,
which is at least an order of magnitude smaller than the intrinsic ionizing photon
flux $N_{\rm Lyc}$, because most of the ionizing photons are absorbed by dust.
The estimate derived from the [Ne{\small II}] line is consistent with this limit
provided that the optical depth in the line be lower than 2.6\,. The modelled
optical depth at the wavelengths of both [Ne{\small II}] and [Ne{\small III}]
is however of the order of 6.5 (Sect.~\ref{model}). This discrepancy calls for
at least one remark: if the starburst is not made of a unique stellar cluster but
occurs simultaneously at several locations (Sect.~\ref{model}), then the
[Ne{\small II}] line emission may arise from the regions of lowest optical depth.

The upper limits obtained on the hydrogen recombination lines \citep{Roussel03}
provide another constraint on the nebular extinction. The requirement that
the recombination rates derived from the 3\,cm, [Ne{\small II}], Br$\gamma$
and Pa$\beta$ measurements be all consistent implies a nebular optical depth
of $0.9 < \tau(12.8\,\mu{\rm m}) < 2.6$, assuming the Galactic center
extinction law of \citet{Moneti01}.

We have assumed that collisional de-excitation is negligible. If this were not
the case, then our estimate of the apparent ionizing photon flux would be underestimated,
and the optical depth in the [Ne{\small II}] line would have to be even lower.
The data do not provide any constraint on the electronic density, but the radiative
transfer model (Sect.~\ref{model}) suggests a central gas density of the order of
$10^4$\,cm$^{-3}$. To determine whether the mid-infrared nebular lines can be
suppressed by collisional de-excitation in NGC\,1377, we computed the critical
densities of the ionic lines that are usually the brightest between 10 and
35\,$\mu$m in star-forming galaxies, and that are tabulated in Table~\ref{tab_lines}.
If the electronic density is below the critical densities, then we expect minimum
line ratios $F_{\rm [SiII]} / F_{\rm [NeII]} \sim 0.5$,
$F_{\rm [SIII]\,33.5\,\mu{\rm m}} / F_{\rm [NeII]} \sim 0.15$
and $F_{\rm [SIII]\,18.7\,\mu{\rm m}} / F_{\rm [NeII]} \sim 0.2$.
These values are derived empirically from the lower bounds observed in nuclei
within the rest of the SINGS galaxy sample~; they correspond to the theoretical
values if $\approx 60$\% of gas-phase silicon is in Si$^+$ assuming that
$\approx 90$\% of silicon atoms are retained in dust grains, and $\approx 5$\%
of sulfur is in S$^{2+}$, with the Si and S solar abundances given by \citet{Grevesse98}.
The upper limits obtained for NGC\,1377 are above these minimum ratios, except
$F_{\rm [SIII]\,18.7\,\mu{\rm m}} / F_{\rm [NeII]} < 0.12$.
Since an optical depth of $\sim 1$ in [Ne{\small II}] (corresponding to an optical
depth of the same order in [Ne{\small III}]) is sufficient to
explain this low ratio, we are unable to set constraints on the electronic density.
Collisional de-excitation may not operate for [Ne{\small II}] and
[Ne{\small III}], since their critical densities are at least ten times higher
than the neutral density estimate from the radiative transfer model. It should
be noted, on one hand, that the peak density could be significantly higher than
our estimate, and, on the other hand, that the electronic density should be much
lower than the neutral density if pressure equilibrium is maintained. Is is thus
unlikely that [Ne{\small II}] and [Ne{\small III}] be collisionally de-excited.

\section{Molecular hydrogen}
\label{gas}

NGC\,1377 is the only member of the SINGS sample in which H$_2$ lines dominate
the line spectrum in the mid-infrared (and also in the near-infrared).
Three rotational transitions of H$_2$ have been confidently detected: the S(1)
and S(2) transitions at high spectral resolution, and the S(3) transition at low
resolution in the SL1 module (Fig.~\ref{fig:h2}). The line fluxes are given in
Table~\ref{tab_lines}. Based on previous observations of rovibrational transitions
in the near-infrared, we had concluded that H$_2$ is collisionally excited
(pure fluorescence being ruled out by the high flux ratio of the v=1-0 S(1)
transition to the v=2-1 S(1) transition), and that slow shocks in the molecular
phase are responsible \citep{Roussel03}. The alternative mechanism, collisions
with hot hydrogen atoms and molecules in dense photodissociation regions
\citep[][and references therein]{Hollenbach97}, seems indeed in conflict with the
expectation that photoelectric heating will be inefficient in NGC\,1377, because
small dust grains may be positively charged due to the very intense radiation
field. This effect was observed by \citet{Malhotra01} for the [C{\small II}] and
[O{\small I}] lines at 157.7\,$\mu$m and 63.2\,$\mu$m, which are among the most
efficient coolants of photodissociation regions. They were not detected by ISO-LWS
observations in NGC\,1377 \citep{Roussel03}.
The power emitted in the sum of the S(1) to S(3) transitions of H$_2$ represents
a fraction $\sim 6 \times 10^{-4}$ of the far-infrared luminosity (40-120\,$\mu$m)~;
and the upper limits on the [C{\small II}] and [O{\small I}] luminosities
respectively $\sim 6 \times 10^{-4}$ and $\sim 9 \times 10^{-4}$.

Using the recent photodissociation region models of \citet{Kaufman06}, we can test
whether this excitation mechanism is efficient enough to account for the observed
brightness of the H$_2$ lines. In these models, the maximum fraction of the
far-ultraviolet radiation (FUV, between 6 and 13.6\,eV) emerging in the rotational
S(1) line is $\sim 10^{-3}$, and it occurs for a radiation field intensity of
$G_0 \sim 200$ times the local value and densities of $n \sim 10^4$\,cm$^{-3}$.
At higher $G_0$ values, the gas heating efficiency declines~; and at higher densities,
the gas cools partially through collisions with dust grains \citep{Hollenbach97}.
The H$_2$ line flux ratios of NGC\,1377 ($F_{\rm S(1)} / F_{\rm S(0)} > 7$,
$F_{\rm S(2)} / F_{\rm S(0)} > 4$, $F_{\rm S(3)} / F_{\rm S(1)} \geq 1$ and
$F_{\rm v=1-0 S(1)} / F_{\rm v=2-1 S(1)} > 30$, these limits accounting for
any amount of extinction) indicate that $G_0 > 10^4$ and
$10^5 < (n / {\rm cm}^{-3}) < 10^7$~; and under these conditions, the maximum
$F_{\rm S(1)} / FUV$ flux ratio is $\sim 4 \times 10^{-4}$. In NGC\,1377,
the 3\,cm upper limit implies that the part of the ionizing photon flux
$N_{\rm Lyc}$ escaping dust absorption is $ < 7.4 \times 10^{52}$\,s$^{-1}$, and
$L_{\rm FUV} \sim 5.1 \times 10^{-18}\,{\rm W} \times (N_{\rm Lyc} / {\rm s}^{-1})$
for a 1\,Myr-old starburst with a Salpeter initial mass function (using the
population synthesis model of \citet{Leitherer99}). Therefore, we expect to
observe a maximum of $\sim 30 \times 10^{-18}$\,W\,m$^{-2}$ in the v=0 S(1)
line.

The observed flux is twice as high. The intrinsic flux can be much higher,
if it is affected by the same optical depth as toward [Ne{\small II}], since gas
heated in photodissociation regions would have to be located very close to the
H{\small II} regions. The minimum optical depth toward [Ne{\small II}]
(Section~\ref{lines}) would correspond to $\tau(17\,\mu{\rm m}) \sim 1.1$.
The intrinsic flux in the H$_2$ v=0 S(1) line would therefore be about six times
higher than the model prediction. This difference is marginal, compared with model
uncertainties, in particular on the H$_2$ collisional de-excitation rates.
However, we note that the Br$\gamma$ to H$_2$ v=1-0 S(1) flux ratio (which is immune
to extinction effects if the emission is nearly cospatial) is lower than 0.1 in
NGC\,1377 \citep{Roussel03}, which is much lower than expected in the case of
photodissociation region heating of H$_2$. For example, the normal starburst
NGC\,1022 has a Br$\gamma$ to H$_2$ v=1-0 S(1) flux ratio of 2.2\,. We thus
argue that shock excitation plays a dominant role in NGC\,1377, although
photodissociation region models are marginally consistent with the present data.

\begin{figure}[!ht]
\hspace*{-1.5cm}
\resizebox{16cm}{!}{\rotatebox{90}{\plotone{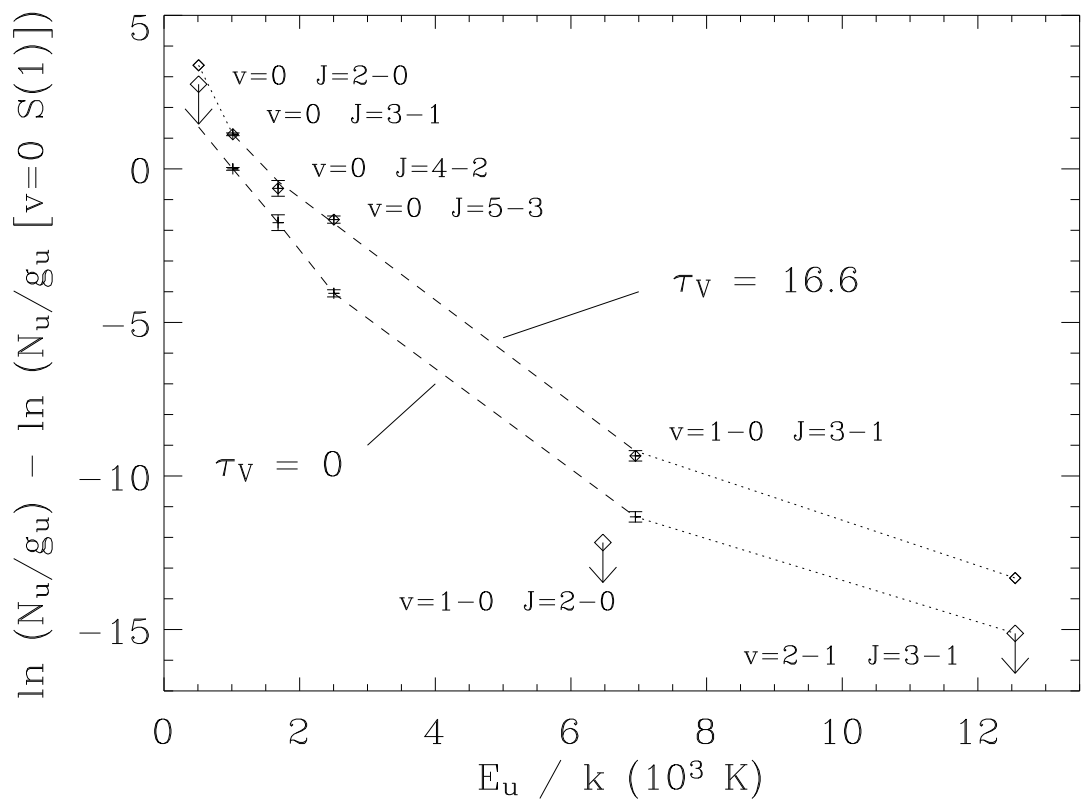}}}
\caption{Excitation diagram for the observed rotational and rovibrational transitions
of H$_2$.
The abscissa is the temperature equivalent to the upper level energy of
the transition, and the ordinate is the logarithm of the number of molecules
in the upper level divided by the statistical weight, normalized by this ratio
for the rotational S(1) line. The lower curve connects the data points uncorrected
for extinction, and the upper curve corresponds to the maximum allowed extinction
(see text), the normalization being done before the extinction correction.
The v=1-0 S(0) upper limit, being inconsistent with the v=1-0 S(1) flux, was
not considered, as it is probably underestimated.
}
\label{fig:diag_h2}
\end{figure}

An excitation diagram combining all the observed transitions to date
(Fig.~\ref{fig:diag_h2}) confirms collisional excitation, and allows us to set
a constraint on the amount of extinction in the H$_2$ lines, as explained below.
The flux of a transition $v \rightarrow v\prime~ J \rightarrow J-2$ is given by
$F = h \nu~ A~ N_{\rm u\,col}~ \Omega / (4 \pi)$, where $N_{\rm u\,col}$ is the
column density of molecules in the upper level, $A$ is the spontaneous emission
probability, $h \nu$ is the transition energy and $\Omega$ is the source solid
angle. We also have
$N_{\rm u} = g_{\rm u}~ N_{\rm tot}~ {\rm exp}(-E_{\rm u}~ /~ (kT))~ /~ Z(T)$, where
$g_{\rm u} = (2 i + 1)~ (2 J + 1)$ is the statistical weight (with the spin number
$i=0$ for even J or para transitions, and $i=1$ for odd J or ortho transitions),
$Z(T) \sim 0.0247~ T~ /~ (1 - exp(-6000\,{\rm K}~ /~ T))$ is the partition function,
and $E_{\rm u}$ is the upper level energy \citep{Herbst96}. Line ratios thus allow
the derivation of an apparent excitation temperature:
\begin{eqnarray*}
kT = (E_{\rm u2} - E_{\rm u1})~ /~ {\rm ln}(N_{\rm u1} / N_{\rm u2} \times g_{\rm u2} / g_{\rm u1}) \\
{\rm with}~~ N_{\rm u1} / N_{\rm u2} = F_1 / F_2 \times A_2 / A_1 \times \lambda_1 / \lambda_2.
\end{eqnarray*}
We used the spontaneous emission probabilities given by \citet{Turner77}.
Since the temperature constrained by the rotation-vibration S(0) lines and by
the rotation-vibration S(1) lines is consistent with the temperature derived
from the pure-rotation transitions, collisional excitation is confirmed.

In order for the excitation temperature to remain monotonic as a function of the
upper level energy ($T$ increasing as $E_{\rm u}$ increases), the equivalent V-band
optical depth has to be lower than 17 ($\tau < 1.1$ at 17\,$\mu$m). This estimate
takes into account the errors on the line fluxes, and assumes that H$_2$ at all
temperatures is affected by a same extinction following the Galactic center
extinction law of \citet{Moneti01}. Since the optical depth in the dust phase
is much higher, most of the detected molecular hydrogen must come from intermediate
layers. For zero extinction, the excitation temperature varies from ($365 \pm 50$)\,K
for the low-energy transitions, to ($610 \pm 25$)\,K between the v=0 S(3) and
v=1-0 S(1) transitions, to $<1500$\,K for the higher-energy transitions~; for
the maximum allowed extinction, it varies between 420\,K and $<1350$\,K.

Computing the mass of warm molecular hydrogen from the flux of the S(1)
transition and the temperature derived from the rotational lines, we obtain
$7.2 \times 10^5$\,M$_{\sun}$ in the hypothesis of zero extinction, and
$1.9 \times 10^6$\,M$_{\sun}$ with $\tau_{\rm V} = 17$. Since we had derived
a total molecular hydrogen mass of the order of $2 \times 10^8$\,M$_{\sun}$
from CO observations \citep{Roussel03}, this implies that less than 1\% of H$_2$
is heated to more than 300\,K. In comparison with the starbursts observed by
\citet{Rigopoulou02}, this fraction is low.
However, in the latter objects, the contribution of photodissociation regions
to the heating of H$_2$ is likely dominant, and the volume fraction of the
molecular gas directly exposed to FUV radiation is probably much larger than
in NGC\,1377.

\section{Alternative hypotheses to a nascent starburst}
\label{seyfert}

Different scenarios to account for the synchrotron deficiency (starburst deficient
in high-mass stars, abnormally weak or strong magnetic field, dominant inverse
Compton losses, extremely high electron opacity) were discussed at length by
\citet{Roussel03}, and it was concluded that the most likely explanation involved
a very young starburst.
The possibility that a substantial population of relativistic electrons exists,
but decays at a higher rate than the type-II supernova rate, in the hypothesis of
inverse Compton losses dominating synchrotron losses, is not formally ruled out,
but seems in conflict with other properties of the starburst. In particular, we did
not detect the [Fe{\small II}] line at 1.644\,$\mu$m, which is a quantitative
tracer of shocks in supernova remnants \citep{Vanzi97}, where cosmic ray electrons
are accelerated~; the line flux ratio of [Fe{\small II}] to H$_2$ at 2.122\,$\mu$m
is below 0.1 in NGC\,1377, whereas it is above 2 for the mature starburst in
NGC\,1022, which we chose as a comparison galaxy (see also \citet{Larkin98} and
\citet{Dale04}). Reconciling these ratios by invoking differential extinction
effects (and assuming cospatial emission) would imply $\tau_{\rm V} > 50$,
whereas we have derived $\tau_{\rm V} < 17$ for H$_2$ (Section~\ref{gas}).
The massive stars in NGC\,1377 are thus unlikely to have reached
the supernova stage. In addition, it is hardly conceivable that a starburst older
than 5\,Myr would still be completely embedded, despite the effects of winds from
massive stars and supernova shock waves on the interstellar medium. On the other
hand, most cosmic rays generated by past star formation episodes would presumably
have propagated over large distances and be located outside the current starburst
region, which is very compact, and thus
outside of reach of the intense radiation field of the starburst. Their absence is
thus more plausibly due to NGC\,1377 being previouly quiescent for a long period,
than to their suppression by inverse Compton losses. Finally, we know of some
ultraluminous galaxies, studied by \citet{Condon91}, that host very compact
starbursts with radiation fields at least as intense as in NGC\,1377~; yet, they
are bright synchrotron emitters, which implies that their magnetic field has been
amplified to the degree necessary to balance inverse Compton losses by synchrotron
losses. It is unclear why magnetic field amplification would be inhibited in
NGC\,1377. In summary, the issue of inverse Compton losses is not settled by the
present data, but we favor the hypothesis of a nascent starburst as much more likely.

Simple energetics arguments also led \citet{Roussel03}
to conclude that it would be exceedingly difficult to account for the observed
properties of NGC\,1377 by accretion onto a massive black hole.
An upper limit on the mass of a black hole at the center of NGC\,1377, of the
order of $2 \times 10^5$\,M$_{\sun}$, is derived consistently both from the lack of
radio continuum emission (using an empirical relation between the minimum radio
power emitted and the black hole mass of radio-quiet active nuclei) and from
the narrow velocity profile of the H$_2$ 2.12\,$\mu$m line (assuming virial
equilibrium). In the unrealistic assumption that the efficiency of dust heating
by the accretion power be 100\%, this power would still have to be above the
Eddington limit for a black hole of such a mass, to account for the observed
infrared luminosity. In addition, substantial free-free
emission would be observed, because a large fraction of the bolometric luminosity
of active nuclei is ionizing~; the expected flux, at wavelengths short enough that
thermal opacity is negligible, is well above the observational upper limit.
A new element is brought by the Spitzer data to constrain the nature of the
power source. The fact that no high-excitation lines are detected makes the
hypothesis of non-stellar activity all the more difficult. In particular, since the
[Ne{\small II}] line is detected, the undetected [Ne{\small III}] line, affected by
almost the same extinction and having a similar critical density,
sets a meaningful constraint on the radiation hardness. Based on the results of
\citet{Sturm02}, pure Seyfert nuclei (i.e. minimally contaminated by star formation)
have $F_{\rm [NeIII]} / F_{\rm [NeII]}$ ratios between 1 and 3. This line flux
ratio is lower than 0.5 in NGC\,1377. We conclude that non-stellar activity
is very unlikely to be responsible for the infrared power of NGC\,1377.

\section{Summary and discussion}
\label{summary}

NGC\,1377 first attracted our attention as being extremely deficient in radio
continuum emission, with respect to its infrared brightness.
Its dust is also heated to very high temperatures, which implies an intense
radiation field. The lack of synchrotron
emission points to the galaxy being previously quiescent for at least 100\,Myr
(which is corroborated by optical spectral synthesis in Section~\ref{optspec}),
and the current star-forming episode being younger than the lifetime of type-II
supernova progenitors \citep{Roussel03}.
Its far-infrared (40--120\,$\mu$m) luminosity is about $4 \times 10^9$\,L$_{\sun}$,
and the bolometric luminosity of the starburst is of the order of
$1.1 \times 10^{10}$\,L$_{\sun}$.
This corresponds to a starburst with a $1.6 \times 10^7$\,M$_{\sun}$ stellar mass,
assuming an age of 1\,Myr and a Salpeter initial mass function between 0.1 and
120\,M$_{\sun}$ (or $6 \times 10^6$\,M$_{\sun}$ for a lower mass cutoff of
1\,M$_{\sun}$), and about $3 \times 10^4$ O stars.
The global cold molecular gas properties are similar to those of starbursts,
with subthermal excitation and a high far-infrared to CO flux ratio.
The molecular gas mass, $\sim 2 \times 10^8$\,M$_{\sun}$, is only 10 to 30 times
higher than the estimated starburst mass, which implies an extremely short
gas exhaustion timescale for NGC\,1377.
The size of the infrared source is of the order of 100\,pc or smaller, and the new
observations reveal that the optical depth in the amorphous silicate absorption bands
is one of the highest observed in extragalactic systems: the apparent optical
depth at 10\,$\mu$m, of the order of 4, is very similar to that of IRAS\,08572+3915,
the most deeply obscured object in the sample of infrared-luminous galaxies of
\citet{Spoon06a}. It should be noted that the modelled optical depth of NGC\,1377
is about five times higher than the apparent depth.
A few absorption bands from dense and warm molecular gas are also present,
although very faint.

The new data confirm that
NGC\,1377 has the characteristics of an extremely young, totally opaque nuclear
starburst. It is in such an early phase of this starburst that the only tracer
of H{\small II} regions yet detected is a very weak [Ne{\small II}] line.
Due to a combination of two effects, suppression of the ionizing photons by dust
and high optical depth in the mid-infrared, the [Ne{\small II}] line probes only
$\sim 0.8$\% of the intrinsic ionizing photon flux of the starburst.
The growth of the H{\small II}
regions is probably still inhibited by the pressure of the dense interstellar medium,
and a large fraction of the massive stars must be still embedded in their parent
molecular cloud. In order to reconcile the presence of very massive stars in nuclear
starbursts with their nebular line ratios typically indicating low excitation
\citep{Thornley00}, \citet{Rigby04} proposed that in such environments, high
pressure causes massive stars to spend a large fraction of their
lifetime in ultracompact H{\small II} regions, where the production of nebular
lines is quelled. NGC\,1377 could represent an extreme example of this phenomenon.
Galactic systems where the dust competes so efficiently with gas for absorption
of the stellar radiation are extremely rare.

If we try to understand the starburst
in NGC\,1377 as a scaled-up version of stellar-size objects (as a mental picture
only), the class of Becklin-Neugebauer (BN) objects may represent the closest
analogs suited to that purpose. BN objects are at the transition between massive
protostars and pre-main-sequence stars exciting ultracompact H{\small II} regions,
after the intense accretion phase has stopped, but before the radiation pressure
has balanced the interstellar medium pressure \citep{Henning86}. The emergent
infrared radiation is strongly self-absorbed like in NGC\,1377, and molecular hydrogen
is excited by shocks, caused by the interaction of stellar outflows with the
molecular envelope. It is unclear whether the same H$_2$ heating mechanism applies
to NGC\,1377. Shocks could be caused by the collective outflows and winds of massive
protostars and young stars~; or they could be caused by a merger event, provided
the disturbances revealed by optical color maps (Sect.~\ref{morph}) may be
interpreted in this fashion. The optical forbidden line spectrum supports the
existence of shocks also in the foreground of the region of activity
(Sect.~\ref{optspec}).

The central neutral density inferred from a radiative transfer model is of the
order of $10^4$ hydrogen nuclei cm$^{-3}$, which is not greater than typical
densities of giant molecular clouds. The estimated starburst mass (see above)
is also very similar to the mass of the super star cluster coincident with
the brightest mid-infrared source in the Antennae galaxies \citep{Gilbert00},
although the latter cluster is more evolved, since it is associated with
bright H{\small II} and photodissociation regions.
In view of these results, we are not able to determine whether the activity
in NGC\,1377 arises from the formation of a single cluster, or from several
isolated pockets. We had been led to the latter hypothesis by the fact that
the molecular gas could be unstable with respect to gravitational instabilities
\citep{Roussel03}, so that star formation could be triggered almost instantaneously
at several locations. The data are however compatible with the formation of
a single stellar cluster in the nucleus. This question should be more easily
addressed for more luminous objects.

\begin{figure}[!ht]
\hspace*{-1.5cm}
\resizebox{16cm}{!}{\rotatebox{90}{\plotone{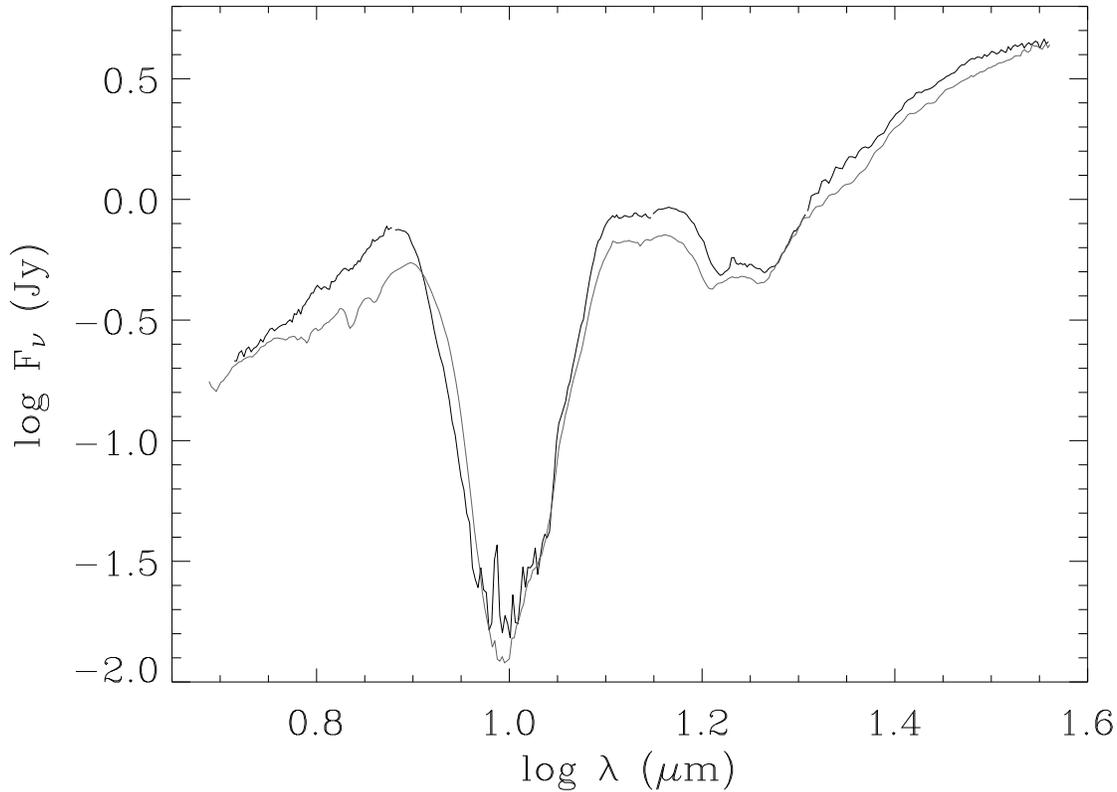}}}
\caption{Comparison of the rest-frame mid-infrared spectra of NGC\,1377
(this study) and IRAS\,08572+3915 \citep{Spoon06a}. The spectrum of NGC\,1377
is shown in black. The spectrum of IRAS\,08572+3915, normalized to that of NGC\,1377
at the long-wavelength end of the LL1 module (36\,$\mu$m), is shown in grey.
The far-infrared (40--120\,$\mu$m) luminosity of IRAS\,08572+3915 is
$\sim 5 \times 10^{11}$\,L$_{\sun}$, about 100 times the far-infrared
luminosity of NGC\,1377, and it has a slightly higher $F_{60\,\mu m}/F_{100\,\mu m}$
ratio of 1.4\,.
}
\label{fig:iras08572}
\end{figure}

At the other end of the mass and luminosity scale, the infrared spectrum of NGC\,1377
is indeed strikingly similar to those of some deeply obscured infrared-luminous
galaxies (Fig.~\ref{fig:iras08572}). NGC\,4418 and IRAS\,08572+3915 for instance
\citep{Spoon01, Spoon06a} share the main spectral characteristics of NGC\,1377.
In addition, these luminous galaxies are part of a statistical sample of nascent
starbursts that we have selected by their deviation from the infrared-radio
correlation and their hot dust. Forthcoming papers
will present and discuss the infrared-radio properties and molecular gas content
of this sample, based on observations with Spitzer, the VLA array and the IRAM 30\,m
telescope. The characteristics of these galaxies will be contrasted with those
of NGC\,1377, serving as an archetype of the nascent starburst class.
Their diverse radio continuum properties suggest that they should sample well
young starbursts at various stages of development \citep{Roussel05}. It will be one
of our aims to sketch an empirical evolutionary sequence from dust-bounded nascent
starbursts such as NGC\,1377 to mature starbursts such as M\,82.

We have previously derived a rate of occurrence of the nascent starburst phenomenon
of $\sim 16$\% among galaxies whose flux density distribution peaks at 60\,$\mu$m
\citep{Roussel03}. This rate is not in conflict with the expected timescales
of star formation. Nascent starbursts, as we have defined them, must fulfill two
conditions: their current star formation episode must be younger than 5\,Myr,
and they must previously have been quiescent for $\sim 100$\,Myr. If we note the
typical duration of compact starbursts $t \times 5$\,Myr, where $t$ is probably
not much larger than unity, then the fraction of starbursts occurring in isolated
mode (as opposed to recurrent mode) would be $0.16 \times t$.

\begin{figure}[!ht]
\hspace*{-1.5cm}
\resizebox{16cm}{!}{\rotatebox{90}{\plotone{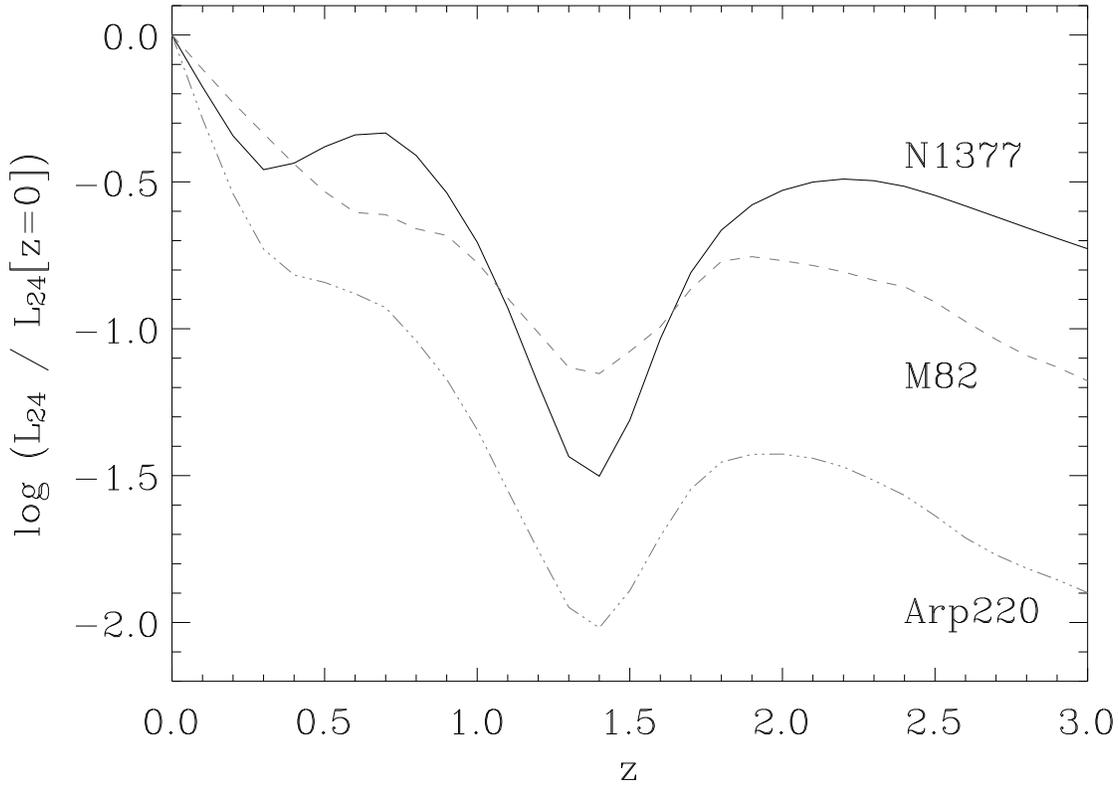}}}
\caption{Evolution of the luminosity in the 24\,$\mu$m MIPS filter as a function
of redshift, for three different mid-infrared spectral energy distributions:
those of NGC\,1377 (solid curve), M\,82 (dashed curve) and Arp\,220 (dot-dashed
curve). M\,82 and Arp\,220 were chosen for this comparison because they are
popular starburst templates.
}
\label{fig:zresponse}
\end{figure}

Starburst tracks currently do not incorporate the pre-H{\small II} nucleus phase.
Likewise, the sets of galactic templates used to interpret the infrared counts
in deep surveys do not include objects such as NGC\,1377.
Since even small changes in the assumed spectrum of aromatic bands in the
mid-infrared yield important differences in the interpretation of the counts,
as emphasized by \citet{Smith06}, the consequences of the existence of luminous
galaxies sharing the properties of NGC\,1377 cannot be neglected.
This is illustrated by Figure~\ref{fig:zresponse}, showing the large differences
in the redshift-luminosity curve for the 24\,$\mu$m MIPS filter, according to
the assumed mid-infrared spectral energy distribution: the curve obtained for
NGC\,1377 is vastly different from those of popular starburst templates.
The nascent starbursts (selected by their deficiency in radio continuum emission,
with respect to the infrared) are rare objects, in agreement with the brevity of
their evolutionary phase. However, they represent a non-negligible fraction of
galaxies with $F_{60\,\mu m} > F_{100\,\mu m}$, whose frequency may be much higher
at earlier epochs, as suggested by the results of \citet{Chapman04}.

\acknowledgements
We wish to thank Marc Sauvage for fruitful discussions, Mark Wolfire for
tirelessly running photodissociation region models to help interpret the data,
and Nancy Silbermann for the timely scheduling of the Spitzer IRS observations.
Support for this work, part of the Spitzer Space Telescope Legacy Science
Program, was provided by NASA through an award issued by the Jet Propulsion
Laboratory, California Institute of Technology under NASA contract 1407.

\clearpage

\begin{deluxetable}{rrr}
\tablecaption{Flux densities of NGC\,1377 in the Spitzer bands (this study~;
see text) and in the ISO and IRAS bands \citep{Roussel03}.
The total photometric uncertainty for the Spitzer measurements is of the order
of 10\% in the IRAC bands, 5\% at 24\,$\mu$m, 10\% at 71\,$\mu$m and 15\% at
156\,$\mu$m~; for the ISOCAM measurements, $\sim 20$\%~; and for the IRAS
measurements, 10\% at 12 and 25\,$\mu$m and 15\% at 60 and 100\,$\mu$m
(IRAS Explanatory Supplement).
\label{tab_flux}
}
\tablehead{
band & nuclear flux & total flux \\
~    & (Jy)         & (Jy) \\
}
\startdata
3.6\,$\mu$m (IRAC)    & 0.0199\tablenotemark{a} & 0.0591 \\
4.5\,$\mu$m (IRAC)    & 0.0520                  & 0.0817 \\
5.7\,$\mu$m (IRAC)    & 0.2715                  & = nuclear \\
6.7\,$\mu$m (ISOCAM)  & ~                       & 0.366 -- 0.391\tablenotemark{b} \\
7.9\,$\mu$m (IRAC)    & 0.4105                  & = nuclear \\
12\,$\mu$m (IRAS)     & ~                       & 0.43 \\
14.3\,$\mu$m (ISOCAM) & ~                       & 0.705 -- 0.721\tablenotemark{b} \\
24\,$\mu$m (MIPS)     & ~                       & 1.740 \\
25\,$\mu$m (IRAS)     & ~                       & 1.83 \\
60\,$\mu$m (IRAS)     & ~                       & 7.01 \\
71\,$\mu$m (MIPS)     & ~                       & 5.64\tablenotemark{c} \\
100\,$\mu$m (IRAS)    & ~                       & 5.97 \\
156\,$\mu$m (MIPS)    & ~                       & 3.19\tablenotemark{c} \\
\enddata
\tablenotetext{a}{If we try to decompose the surface brightness profile in the
3.6\,$\mu$m band to estimate the central brightness of the disk, and adjust
a point spread function to the residual central brightness above this level,
we obtain an estimate of 0.0162 Jy for the nuclear flux, i.e. 10\% lower.}
\tablenotetext{b}{The two flux values for each ISOCAM band correspond to
measurements before and after correction for a slight saturation of the nuclear
pixels.}
\tablenotetext{c}{The MIPS 71 and 156\,$\mu$m fluxes include the latest flux
calibration adjustments ($+ 11$\% and $+ 6$\% respectively, from version S12
to version S13 of the Spitzer Science Center pipeline).
}
\end{deluxetable}

\begin{deluxetable}{lrlrr}
\tablecaption{
Line fluxes and upper limits and critical densities of the forbidden lines\tablenotemark{a}.
All lines were measured at high spectral resolution, except H$_2$ v=0 S(3) which is
outside the wavelength coverage of the high-resolution modules.
The uncertainties on the critical densities reflect a range of electronic temperature
between $5 \times 10^3$\,K and $10^4$\,K.
\label{tab_lines}
}
\tablehead{
line & rest wavelength & spectral order & flux                      & critical density \\
~    & ($\mu$m)        & ~              & ($10^{-18}$\,W\,m$^{-2}$) & (cm$^{-3}$) \\
}
\startdata
H$_2$ v=0 S(0)       & 28.219 & LH\,14 & $< 9.15$          & ~ \\
H$_2$ v=0 S(1)       & 17.035 & SH\,12 & $65.43 \pm 2.94$  & ~ \\
H$_2$ v=0 S(2)       & 12.279 & SH\,17 & $39.14 \pm 10.03$ & ~ \\
H$_2$ v=0 S(3)       &  9.665 & SL\,1  & $65.10 \pm 7.64$  & ~ \\
$[$Si{\small II}$]$  & 34.814 & LH\,11 & $< 11.93$         & $(1.5 \pm 0.3) \times 10^3$ \\
$[$S{\small III}$]$  & 33.480 & LH\,12 & $< 7.20$          & $(6.1 \pm 0.9) \times 10^3$ \\
$[$S{\small III}$]$  & 18.713 & SH\,11 & $< 2.00$          & $(2.0 \pm 0.3) \times 10^4$ \\
$[$Ne{\small III}$]$ & 15.555 & SH\,13 & $< 8.97$          & $(2.4 \pm 0.4) \times 10^5$ \\
$[$Ne{\small II}$]$  & 12.814 & SH\,16 & $17.38 \pm 2.13$  & $(6.0 \pm 1.0) \times 10^5$ \\
\enddata
\tablenotetext{a}{
The critical densities are computed as:
$N_{\rm crit} = A~ /~ [C~ (T_{\rm e} / 1\,{\rm K})^{-0.5}~ \Omega(T_{\rm e})~ g_{\rm u}^{-1}]$,
where $A$ is the transition probability, taken in the NIST atomic spectra
database ({\it http://physics.nist.gov/PhysRefData/ASD}),
$C = 8.6287 \times 10^{-6}$\,cm$^3$\,s$^{-1}$, and $g_{\rm u}$
is the statistical weight of the upper level. We used collision strengths from
\citet{Galavis95} for the [S{\small III}] lines and \citet{Dufton91} for the
[Si{\small II}] line.
}
\end{deluxetable}

\clearpage

\clearpage

\clearpage

\end{document}